\documentclass[twocolumn,showpacs,preprintnumbers,amsmath,amssymb]{revtex4}
\usepackage{pifont}

\usepackage{graphicx}
\usepackage{dcolumn}
\usepackage{bm}
\usepackage[colorlinks,linkcolor=blue,anchorcolor=blue,citecolor=blue]{hyperref}
\usepackage{lineno} 
\usepackage{epstopdf}
\usepackage{amssymb}
\usepackage{amsmath}
\usepackage{latexsym}
\usepackage{mathrsfs}

\begin{document}
\preprint{}

\title{L\'{e}vy walk dynamics in mixed potentials from the perspective of random walk theory}

\author{Tian Zhou$^{1}$}
\author{Pengbo Xu$^{1}$}
\author{Weihua Deng$^{1}$}
\affiliation{$^1$School of Mathematics and Statistics, Gansu Key Laboratory of Applied Mathematics and Complex Systems, Lanzhou University, Lanzhou 730000, P.R. China}



\begin{abstract}
L\'evy walk process is one of the most effective models to describe superdiffusion, which underlies some important movement patterns and has been widely observed in the micro  and macro dynamics. From the perspective of random walk theory, here we investigate the dynamics of L\'evy walks under the influences of the constant force field and the one combined with harmonic potential. Utilizing Hermite polynomial approximation to deal with the spatiotemporally coupled analysis challenges, some striking features are detected, including non Gaussian  stationary distribution, faster diffusion, and still strongly anomalous diffusion, etc.


\end{abstract}

\maketitle

\section{Introduction}

The diffusion processes are usually classified according to the relation between the mean squared displacement (MSD) $\langle x^2(t)\rangle=\int x^2P(x,t)dx$ and time $t$, which more often appears as  $\langle x^2(t)\rangle\sim t^\alpha$. To be more specific, the process is called anomalous diffusion when $\alpha\neq 1$ \cite{BG1990,SLLS2004,MJCB2014}, otherwise it is called normal diffusion such as Brownian motion \cite{NelsonE}. Specifically, subdiffusion with $0<\alpha<1$ can be observed in a wide range of areas \cite{SM1975,SW2009,JLOM2013,WSTK2011,JMJM2012,WST2010,TBKK2013}, and superdiffusions with $\alpha>1$ are also ubiquitous \cite{CGE2000,BS2002,RJLM2015}. One of the most popular models to describe anomalous diffusion is continuous time random walk (CTRW), which consists of two series of independent identically distributed (i.i.d.) random variables: one is waiting time $\tau$ following the probability density function (PDF) $\phi(\tau)$, and the other one is jump length denoted as $l$ with PDF $\rho(l)$; additionally for CTRW process we assume space and time are independent with each other \cite{s11,M1969,MK2000}. The corresponding subdiffusion process then can be considered as a scaling limit process of CTRW with infinite $\langle \tau\rangle$ and finite $\langle l^2\rangle$. While for superdiffusion process, we usually consider $\tau$ having finite average and $\rho(l)\sim1/|l|^{1+\mu}$ with $0<\mu<2$ so that $\langle l^2\rangle$ diverges, and the corresponding model is called L\'evy flight \cite{F1994a,F1994b,MK2000}. In fact, if one directly calculates MSD of L\'evy flight process, the result obtained is always infinite. However, people usually calculate the fractional moment $\langle\vert x\vert^q\rangle$ instead, and then take a pseudo limit as $q\to 2$, the exponent of which indicates that the process belongs to superdiffusion \cite{MK2000}.

Based on random walk theory, another well-developed and popular model to describe anomalous diffusion is L\'evy walk, which has a coupled space and time through finite propagation speed \cite{s13}. The MSD of L\'evy walk is bounded, because of the finite propagation speed, which in some sense makes it more physical in practical applications  \cite{CDP2005,BFK2000,LTMJ2008,VGMN2019}. The traditional L\'evy walk process consists of a series of i.i.d. random variables known as running time $\tau$, and a constant value of velocity $v_0$, therefore the movement of each finished step is $\pm v_0\tau$ \cite{s13}. Besides, L\'evy walk has many other generalizations: in \cite{ZSS2008} the velocity of L\'evy walk process is considered as a random variable, \cite{ZXD2020} discusses L\'evy walk process with stochastic resetting to the origin, and \cite{Pengbo} generalizes the range of anomalous diffusion described by L\'evy walk from superdiffusion and normal diffusion to subdiffusion, etc.

Most of the time, the particles are moving in some (one or several) external potentials, e.g., the Earth's gravitational potential, harmonic oscillator potential,  optical lattice potential, elliptic function potential, and double well potential, etc.
The CTRW processes moving in external potential are considered in \cite{MBK1999,JMF1999,CKGM2003}, and the MSD of L\'evy flight process in harmonic potential is discovered to be divergent,  although this process has a stationary state \cite{JMF1999}. Doing the analysis of the L\'evy walk process in external potential is not an easy issue, since the Laplace and Fourier transform methods lose their advantages in solving the problem with coupled space and time. The L\'evy walk with constant external drift has been discussed in \cite{SIM2003}, and the generalized Kramers-Fokker-Planck equation which governs the PDF of L\'evy walk in arbitrary external potential is given in \cite{FJBE2006a,FJB2006b}; although the equation has been given, it seems that the concrete properties are still too hard to be directly obtained from Kramers-Fokker-Planck equation. In \cite{chen}, based on the Langevin picture of L\'evy walk, the influence of constant external force is considered. More recently, \cite{Pengbo} establishes a completely new method to solve the problem of L\'evy walk by introducing Hermite polynomials, which is a compensate method to the traditional integral transform one. Furthermore, in \cite{Pengbo1} the L\'evy walk process in harmonic potential has been detailedly discussed. In this paper, we will continue to discuss the L\'evy walk problem in constant external force from the direct perspective of random walk and some important statistical quantities are calculated;  then the L\'evy walk particles in constant force field combined with harmonic potential are also fully discussed, and the similarities and the differences between this mixed potential and pure harmonic potential are carefully studied.

This paper is organized as follows. In Sec. \ref{sec 2}, we introduce the L\'{e}vy walk model in constant external force field, analyze its fractional moment, and calculate the corresponding average displacement, MSD, and variance by the approach of Hermite polynomial expansion. In Sec. \ref{sec 3}, we turn to discuss L\'{e}vy walk model in the combined external harmonic potential and constant force field; again some representative statistical quantities are calculated, and we analyze the stationary distribution as well as the relaxation dynamics. Finally, we conclude the paper with some discussions.

%

\section{L\'evy walk in a constant force field}\label{sec 2}

\subsection{Introduction of the model}\label{sec2_a}
First we consider the L\'evy walk particles in one dimensional space with mass $M$ move in a constant force field $F=M a$, where $a$ represents constant acceleration. According to the classical L\'evy walk model \cite{s13}, we assume the initial velocity of each step is $\pm v_0$ with probability of $1/2$ to choose either direction. Besides we denote $\{t_j\}_{j=1,\ldots,n}$ as the time when the $j$-th renewal event just finishes, and assume the duration time between two adjacent renewal events $\tau:=t_{j}-t_{j-1}$ obeys running time PDF $\phi(\tau)$. Therefore in this case for each $j=1,\ldots,n$ and $t_j\leq t$, the dynamic between $j$-th and $(j+1)$-th renewal events satisfies the equation:
\begin{equation}\label{1.1}
\left\{
\begin{split}
	&\frac{d^2 x_{t_{j}+\tau'}}{d {\tau'}^2}=a,\quad \text{for}~\tau' \in (0, \{t_{j+1}\wedge t\}-t_{j}],\\
	&\left. \frac{d x_{t_{j}+\tau'}}{d {\tau'}}\right|_{\tau'=0}=\pm v_0.
\end{split}
\right.
\end{equation}
Then the initial velocity $\pm v_0$ gives the corresponding solution to \eqref{1.1}, which is $x_{t_{j}+\tau'}=\frac{1}{2} a \tau'^2\pm v_0 \tau'+x_{t_{j}}$ with probability $1/2$ to choose each of them.


Similar to the discussions of ordinary L\'evy walk model in \cite{ZSS2008}, the PDF of the particle just having arrived at the position $x_t$ at time $t$ is denoted as $q(x_t,t)$. Combining with the above dynamics, we have
\begin{widetext}
	\begin{equation}\label{1.2}
\begin{split}
    q(x_t,t)=&\frac{1}{2}\int_{-\infty}^{\infty}\int_{0}^{t}\delta\left(x_t-\frac{1}{2}a \tau^2-v_0 \tau-x_{t-\tau}\right)\phi(\tau)q(x_{t-\tau},t-\tau)d\tau dx_{t-\tau}\\
      &+\frac{1}{2}\int_{-\infty}^{\infty}\int_{0}^{t}\phi(\tau)q(x_{t-\tau},t-\tau)\delta\left(x_t-\frac{1}{2}a \tau^2+v_0 \tau-x_{t-\tau}\right)d\tau dx_{t-\tau} +P_0(x)\delta(t),
\end{split}
\end{equation}
where $P_0(x)$ is the initial distribution. Here it can be noted that the first two terms on the right hand side of \eqref{1.2} represent the probability of particles transiting from position $x_{t-\tau}$ at time $t-\tau$ to $x_t$ after duration time $\tau$.
Besides, the PDF of L\'evy particles locating at position $x_t$ at time $t$ denoted as $P(x_t,t)$ satisfies
\begin{equation}\label{1.3}
  \begin{split}
    P(x_t,t)=&\frac{1}{2}\int_{-\infty}^{\infty}\int_{0}^{t}\delta\left(x_t-\frac{1}{2}a \tau^2-v_0 \tau-x_{t-\tau}\right)\Phi(\tau)q(x_{t-\tau},t-\tau)d\tau dx_{t-\tau}\\
      &+\frac{1}{2}\int_{-\infty}^{\infty}\int_{0}^{t}\delta\left(x_t-\frac{1}{2}a \tau^2+v_0 \tau-x_{t-\tau}\right)\Phi(\tau)q(x_{t-\tau},t-\tau)d\tau dx_{t-\tau},
\end{split}
\end{equation}
\end{widetext}
%
where the survival probability $\Phi(\tau)$ is defined as
\begin{equation}\label{survival_prob}
  \Phi(\tau)=\int_{\tau}^{\infty}\phi(\tau')d\tau'.
\end{equation}

According to the property of Dirac-delta function, \eqref{1.2} and \eqref{1.3} can be respectively simplified as
\begin{equation}\label{1.4}
 \begin{split}
    q(x_t,t)&=\frac{1}{2}\int_{0}^{t}\phi(\tau)q(x_t-\frac{1}{2}a \tau^2-v_0 \tau,t-\tau)d\tau\\
    &
      +\frac{1}{2}\int_{0}^{t}\phi(\tau)q(x_t-\frac{1}{2}a \tau^2+v_0 \tau,t-\tau)d\tau\\
     &+P_0(x)\delta(t),
\end{split}
\end{equation}
and
\begin{equation}\label{1.5}
  \begin{split}
    P(x_t,t)&=\frac{1}{2}\int_{0}^{t}\Phi(\tau)q(x_t-\frac{1}{2}a \tau^2-v_0 \tau,t-\tau)d\tau\\
    &
      +\frac{1}{2}\int_{0}^{t}\Phi(\tau)q(x_t-\frac{1}{2}a \tau^2+v_0 \tau,t-\tau)d\tau.
\end{split}
\end{equation}

In order to give the explicit form of $P(x_t,t)$ and further calculate some statistical quantities such as MSD, one of the most widely used methods is Fourier transform
$\tilde{f}(k)=\mathscr{F}_{x\rightarrow k}\{f(x)\}=\int_{-\infty}^{\infty}e^{-i k x}f(x)dx$ or Laplace transform $\hat{g}(s)=\mathscr{L}_{t\rightarrow s}\{g(t)\}=\int_{0}^{\infty} e^{-s t}g(t)dt$. After some calculations, from \eqref{1.4} and \eqref{1.5}, the PDF of L\'{e}vy walk particle in a constant force field can be represented as
\begin{equation}\label{1.9}
  \widehat{\widetilde{P}}(k,s)=\frac{\widetilde{P}_0(k)\mathscr{L}_{\tau\rightarrow s}\big\{\exp\left(- \frac{i k a\tau^2}{2}\right)\cos(k v_0 \tau)\Phi(\tau)\big\}}{1-\mathscr{L}_{\tau\rightarrow s}\big\{\exp\left(- \frac{i k a\tau^2}{2}\right)\cos(k v_0 \tau)\phi(\tau)\big\}}.
\end{equation}
From \eqref{1.9}, one can verify the normalization of $P(x,t)$, which is
\begin{equation*}
\begin{split}
	\int_{-\infty}^{\infty} \mathscr{L}^{-1}_{s\to t}\left\{\widehat{P}(x,s)\right\}dx &= \mathscr{L}^{-1}_{s\to t}\left\{\widehat{\widetilde{P}}(k,s)|_{k=0}\right\}\\
	&=\mathscr{L}^{-1}_{s\to t}\left\{1/s\right\}=1.	
\end{split}
\end{equation*}
Although the explicit form of $P(x_t,t)$ in Fourier-Laplace space is given in \eqref{1.9},
since the corresponding Laplace transform is hard to calculate, there are still some troubles in further calculating MSD or other quantities, which motivates us to utilize the Hermite polynomial approximation approach to solve the problem.


\subsection{Hermite polynomial approximation to L\'{e}vy walk in constant force field}

In this subsection, we utilize the Hermite polynomials to approach the PDF of L\'{e}vy walk in a constant force field. The Hermite polynomials form an orthogonal basis of the Hilbert space with the inner product $\langle f,g \rangle=\int_{-\infty}^{\infty} f(x)\bar{g}(x)e^{-x^2} dx$ \cite{hermit_intro}. According to the complete orthogonal system, we assume that in Hilbert space $q(x,t)$ and $P(x,t)$ can be, respectively, represented as
\begin{align}
	q(x,t)&=\sum_{n=0}^{\infty} H_n(x) T_n(t) e^{-x^2},\label{1.10}\\
	P(x,t)&=\sum_{n=0}^{\infty} H_n(x) R_n(t) e^{-x^2},\label{1.11}
\end{align}
where $H_n(x), n=0,1,\cdots,$ represents the Hermite polynomials, $\{T_n(t)\}$ and $\{R_n(t)\}$ are series of functions with respect to $t$ to be determined.

First inserting \eqref{1.10} into \eqref{1.4} leads to
\begin{widetext}
\begin{equation}\label{1.12}
  \begin{split}
     \sum_{n=0}^{\infty} H_n(x)T_n(t) e^{-x^2}&=\frac{1}{2}\int_{0}^{t}\sum_{n=0}^{\infty} H_n\left(x-\frac{1}{2}a \tau^2-v_0 \tau\right)
      T_n(t-\tau) \exp\left[-\left(x-\frac{1}{2}a \tau^2-v_0 \tau\right)^2\right] \phi(\tau) d\tau\\
      & +\frac{1}{2}\int_{0}^{t}\sum_{n=0}^{\infty} H_n\left(x-\frac{1}{2}a \tau^2+v_0 \tau\right)
      T_n(t-\tau) \exp\left[-\left(x-\frac{1}{2}a \tau^2+v_0 \tau\right)^2\right] \phi(\tau) d\tau
       +P_0(x)\delta(t).
  \end{split}
\end{equation}
Multiplying $H_m(x)$, $m=0,1,2,\cdots,$ on both side of \eqref{1.12}, integrating $x$ over $(-\infty,+\infty)$, and utilizing the properties of Hermite polynomials \eqref{a2} and \eqref{a3} in Appendix \ref{Appen_A}, we have
\begin{equation}\label{1.16}
 \sqrt{\pi} 2^m m!T_m(t)= \frac{1}{2}\sum_{k=0}^{m}\frac{m!}{k!(m-k)!}\int_{0}^{t}\sqrt{\pi}2^k k!\left[(a \tau^2+2 v_0 \tau)^{m-k}+(a \tau^2-2 v_0 \tau)^{m-k}\right]T_k(t-\tau)\phi(\tau)d\tau+\delta(t)H_m(0),
\end{equation}
where we choose the initial distribution of particles as Dirac-delta function, i.e., $P_0(x)=\delta(x)$. Finally the iteration relation of ${\hat{T}_m(s)}$ can be obtained by applying Laplace transform on \eqref{1.16} from $t$ to $s$,
\begin{equation}\label{1.17}
 \sqrt{\pi} 2^m m!\widehat{T}_m(s)= \frac{1}{2}\sum_{k=0}^{m}\frac{2^k \sqrt{\pi} m! }{(m-k)!}\mathscr{L}_{\tau\rightarrow s}\Big\{\big[(a \tau^2+2 v_0 \tau)^{m-k}+(a \tau^2-2 v_0 \tau)^{m-k}\big]\phi(\tau)\Big\}\widehat{T}_k(s)+H_m(0).
\end{equation}

Similarly, by substituting \eqref{1.11} into \eqref{1.5}, we find the relation between $\widehat{R}_m(s)$ and $\widehat{T}_m(s)$ in Laplace space
\begin{equation}\label{1.23}
 \sqrt{\pi} 2^m m!\widehat{R}_m(s)= \frac{1}{2}\sum_{k=0}^{m}\frac{2^k \sqrt{\pi} m! }{(m-k)!}\mathscr{L}_{\tau\rightarrow s}\Big\{\big[(a \tau^2+2 v_0 \tau)^{m-k}+(a \tau^2-2 v_0 \tau)^{m-k}\big]\Phi(\tau)\Big\}\widehat{T}_k(s).
\end{equation}
\end{widetext}

In order to obtain the form of $P(x,t)$ as shown in \eqref{1.11}, it is necessary to first calculate the expressions of $\widehat{T}_m(s)$ from the iteration relation \eqref{1.17} for each $m=0,1,\ldots$, and further $\widehat{R}_m(s)$ can be obtained according to \eqref{1.23}. Moreover, it can be noted that to calculate $N$-th moment with $N$ being a non-negative integer one only needs to calculate the first $N$ terms of $\widehat{T}_m(s)$ and $\widehat{R}_m(s)$ instead of infinite terms, bringing us the possible method to calculate the average, MSD, and other higher moment of L\'evy walk process in constant force field.


Now we verify the normalization of $P(x,t)$ given in \eqref{1.11}.
From the definition of Hermite polynomials in \eqref{a1},
the PDF can be rewritten in the form
\begin{equation}\label{1.26.1}
  P(x,t)=\sum_{n=0}^{\infty}(-1)^n \frac{d^n}{d x^n} e^{-x^2} R_n(t),
\end{equation}
therefore applying the Fourier transform $x\to k$ and Laplace transform $t\to s$ on \eqref{1.26.1} leads to
\begin{equation}\label{1.27}
  \widehat{\widetilde{P}}(k,s)=\sum_{n=0}^{\infty} \sqrt{\pi}(-i k)^n e^{-\frac{k^2}{4}} \widehat{R}_n(s).
\end{equation}
Checking the normalization is equivalent to obtain $\widehat{\widetilde{P}}(k=0,s)=1/s$; in fact, from \eqref{1.17} taking $m=0$ and noticing $H_0(x)=1$, we have
\begin{equation}\label{1.18}
  \widehat{T}_0(s)=\frac{1}{\sqrt{\pi}\big(1-\hat{\phi}(s)\big)},
\end{equation}
while from \eqref{1.23} and \eqref{1.27} one can obtain
\begin{align}
	&\widehat{R}_0(s)=\frac{1-\hat{\phi}(s)}{s}\widehat{T}_0(s)=\frac{1}{\sqrt{\pi}s}, \label{1.24}\\
	&\widehat{\widetilde{P}}(k=0,s)=\sqrt{\pi} \widehat{R}_0(s)=1/s.
\end{align}

%
\subsection{Properties of L\'evy walk in constant force field: average displacement, MSD, and strongly anomalous diffusion}

We first consider the average displacement and MSD of L\'evy walk particles for some specific and representative $\phi(\tau)$ moving in a constant force field. From the well-known relation
\begin{equation}\label{mmt}
	\langle x^m(t)\rangle=(i)^m \left. \frac{\partial^m}{\partial k^m} \widetilde{P}(k,t)\right|_{k=0}
\end{equation}
 and \eqref{1.27}, the first two moments of the process in Laplace space can be, respectively, represented as
\begin{equation}\label{firstmoment}
   \langle \hat{x}(s)\rangle =i \left.\frac{\partial}{\partial k} \widehat{\widetilde{P}}(k,s)\right|_{k=0}=\sqrt{\pi} \widehat{R}_1(s),
\end{equation}
and
\begin{equation}\label{msd}
\begin{split}
   \langle\hat{x}^2(s)\rangle & =- \frac{\partial^2}{\partial k^2} \widehat{\widetilde{P}}(k,s)|_{k=0}\\
   &=\frac{\sqrt{\pi}}{2} \widehat{R}_0(s)+2\sqrt{\pi} \widehat{R}_2(s).
\end{split}
\end{equation}
Therefore, in order to calculate the Laplace form of average displacement $\langle \hat{x}(s)\rangle$ and MSD $\langle\hat{x}^2(s)\rangle$, we need to further get $\widehat{R}_1(s)$ and $\widehat{R}_2(s)$, which is necessary to first calculate $\widehat{T}_1(s)$ and $\widehat{T}_2(s)$ based on \eqref{1.18}. Taking $m=1$, from the iteration relation \eqref{1.17} and the value of Hermite polynomials at the origin \eqref{a5}, it can be obtained that
\begin{equation}\label{1..18}
\begin{split}
   2  \sqrt{\pi}\widehat{T}_1(s) =\sqrt{\pi}\mathscr{L}_{\tau\rightarrow s}\big\{ a \tau^2\phi(\tau)\big\}\widehat{T}_0(s)+2\sqrt{\pi}\hat{\phi}(s)\widehat{T}_1(s),
\end{split}
\end{equation}
combining with \eqref{1.18} leads to
\begin{equation}\label{1.19}
 \hat{T}_1(s)=\frac{a \hat{\phi}''(s)}{2\sqrt{\pi}\big(1-\hat{\phi}(s)\big)^2}.
\end{equation}
Further from \eqref{1.17}, the equation of $\widehat{T}_2(s)$ is obtained as
\begin{equation}\label{1.20}
  \begin{split}
&\sqrt{\pi}2^3\widehat{T}_2(s)=2\sqrt{\pi} \mathscr{L}_{\tau\rightarrow s}\big\{2a \tau^2\phi(\tau)\big\}\widehat{T}_1(s)+\\
       &\frac{1}{2}\sqrt{\pi}\widehat{T}_0(s)\mathscr{L}_{\tau\rightarrow s}\Big\{\big[(a \tau^2+2 v_0 \tau)^{2}+(a \tau^2-2 v_0 \tau)^{2}\big]\phi(\tau)\Big\}\\
       &+2^2 \sqrt{\pi}\mathscr{L}_{\tau\rightarrow s}\big\{2\phi(\tau)\big\}\widehat{T}_2(s)+H_2(0),
  \end{split}
\end{equation}
the solution of which is
\begin{equation}\label{1.21}
\begin{split}
   \hat{T}_2(s)=&\frac{2+2 \hat{\phi}(s)^2-4 v_0^2\hat{\phi}''(s)-2 a^2\hat{\phi}''(s)^2-a^2 \hat{\phi}^{(4)}(s)}{8 \sqrt{\pi} \big(\hat{\phi}(s)-1\big)^3} \\
     & +\frac{\hat{\phi}(s)\big(-4+4 v_0^2 \hat{\phi}''(s)+a^2 \hat{\phi}^{(4)}(s)\big)}{8 \sqrt{\pi} \big(\hat{\phi}(s)-1\big)^3}.
\end{split}
\end{equation}


Similarly from \eqref{1.23} and \eqref{1.18}, \eqref{1.19}, \eqref{1.21}, the forms of $\widehat{R}_1(s)$ and $\widehat{R}_2(s)$ yield
\begin{align}
	&\widehat{R}_1(s)=\frac{a\Big[\widehat{\Phi}(s)\widehat{\phi}''(s)-\big(-1+\hat{\phi}(s)\big)\hat{\Phi}''(s)\Big]}{2\sqrt{\pi}\big(\hat{\phi}(s)-1\big)^2 }\label{1.25},\\
	&\widehat{R}_2(s)=\frac{a^2\hat{\phi}''(s)\widehat{\Phi}''(s) }{4\sqrt{\pi}\big(1-\hat{\phi}(s)\big)^2}+\frac{4 v_0^2 \widehat{\Phi}''(s)+a^2 \widehat{\Phi}^{(4)}(s)}{8\sqrt{\pi}\big(1-\hat{\phi}(s)\big)}\nonumber \\
    &~~+\widehat{\Phi}(s)\frac{2+2\hat{\phi}(s)^2-4 v_0^2\hat{\phi}''(s)-2 a^2 \hat{\phi}''(s)^2-a^2 \hat{\phi}^{(4)}(s) }{8\sqrt{\pi}\big(\hat{\phi}(s)-1\big)^3 }\nonumber \\
    &~~~+\widehat{\Phi}(s) \frac{\hat{\phi}(s)\big(-4+4 v_0^2\hat{\phi}''(s)+a^2 \hat{\phi}^{(4)}(s) \big)}{8\sqrt{\pi}\big(\hat{\phi}(s)-1\big)^3}.\label{1.26}
\end{align}
The first two moments of L\'{e}vy walk in a constant force field can also be obtained by, respectively, combining \eqref{firstmoment} and \eqref{msd} with \eqref{1.24}, \eqref{1.25}, and \eqref{1.26}. The average displacement behaves as
\begin{equation}\label{1.28}
   \langle \hat{x}(s)\rangle=\frac{a\Big(\widehat{\Phi}(s)\hat{\phi}''(s)-\big(\hat{\phi}(s)-1\big)\widehat{\Phi}''(s)\Big)}{2\big(\hat{\phi}(s)-1\big)^2 },
\end{equation}
and the MSD can be expressed as
\begin{equation}\label{1.29}
\begin{split}
   \langle\hat{x}^2(s)\rangle=&\frac{1}{2 s}+\frac{a^2\hat{\phi}''(s)\widehat{\Phi}''(s) }{2\big(1-\hat{\phi}(s)\big)^2}+\frac{4 v_0^2 \widehat{\Phi}''(s)+a^2 \widehat{\Phi}^{(4)}(s)}{4\big(1-\hat{\phi}(s)\big)}\\
 &+\widehat{\Phi}(s) \frac{\hat{\phi}(s)\big(-4+4 v_0^2\hat{\phi}''(s)+a^2 \hat{\phi}^{(4)}(s) \big)}{4\big(\hat{\phi}(s)-1\big)^3}\\
  &+\widehat{\Phi}(s)\frac{2+2\hat{\phi}(s)^2-4 v_0^2\hat{\phi}''(s)}{4\big(\hat{\phi}(s)-1\big)^3 } \\
  &-\widehat{\Phi}(s)\frac{2 a^2 \hat{\phi}''(s)^2+a^2 \hat{\phi}^{(4)}(s)}{4\big(\hat{\phi}(s)-1\big)^3 }.
\end{split}
\end{equation}

\begin{figure}[htbp]
  \centering
  \includegraphics[width=8cm]{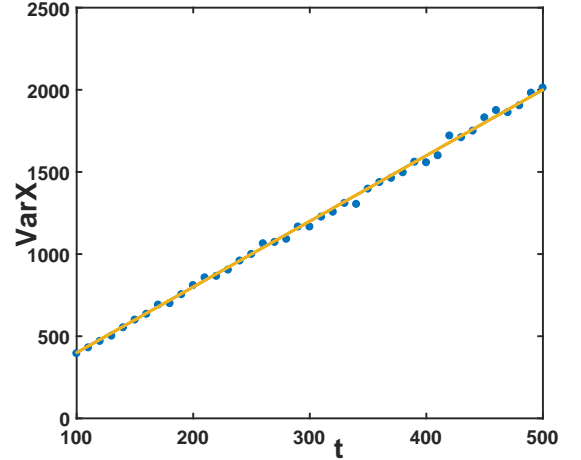}
  \caption{Variance of L\'evy walk process in constant force field with exponentially distributed running time. The parameters are $v_0=F=1, x_0=0$, and $\lambda=1$, and the dots are obtained by averaging over $10^4$ realizations.}
  \label{CFVarexpalpha1}
\end{figure}

In particular, we consider the case that the running time follows exponential distribution, i.e., $\phi(\tau)=\lambda e^{-\lambda\tau}$. Then the corresponding Laplace transform is $\hat{\phi}(s)=\frac{\lambda}{\lambda+s}$ and in this case the survival probability in Laplace space is $\widehat{\Phi}(s)=\frac{1}{\lambda+s}$. Finally from \eqref{1.28} and \eqref{1.29}, the average displacement and MSD in Laplace space can be given as
\begin{equation}\label{1.30}
\begin{split}
   \langle \hat{x}(s)\rangle &=\frac{a}{s^2 (\lambda+s)},  \\
    \langle\hat{x}^2(s)\rangle & =\frac{2 (a^2 (\lambda+3 s)+s (\lambda+s)^2 v_0^2 s)}{s^3 (\lambda+s)^3}.
\end{split}
\end{equation}
Then for sufficiently long time $t$ (small $s$), there are the following asymptotic behaviors after neglecting the high order small terms of $s$ and applying inverse Laplace transform from $s$ to $t$,
\begin{equation}\label{1.31}
 \langle x(t)\rangle\sim\frac{a}{\lambda} t,~~~~~ \langle x^2(t)\rangle\sim\frac{a^2}{\lambda^2}t^2.
\end{equation}
\begin{figure}[htbp]
  \centering
  \includegraphics[scale=0.19]{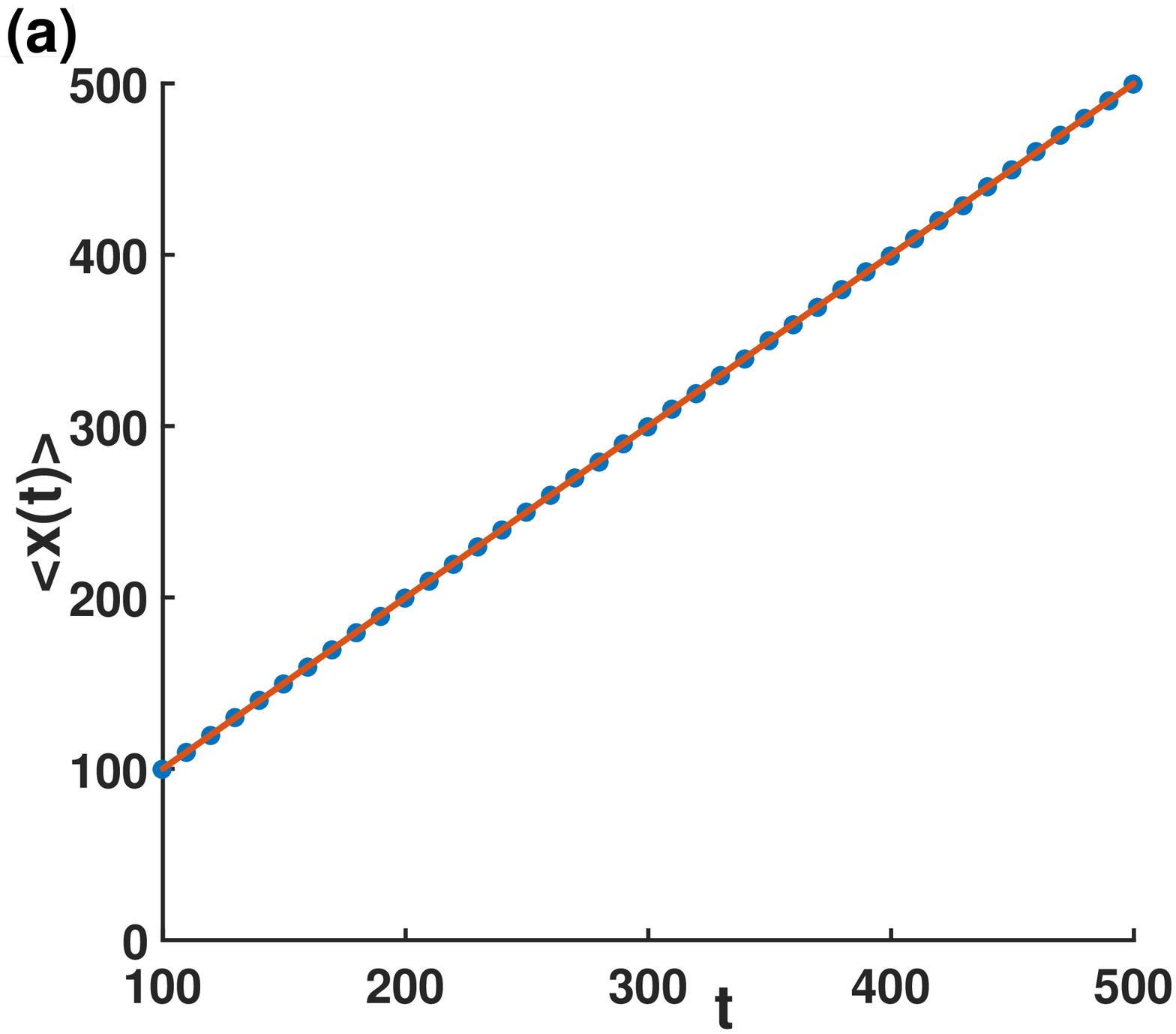}
  \includegraphics[scale=0.19]{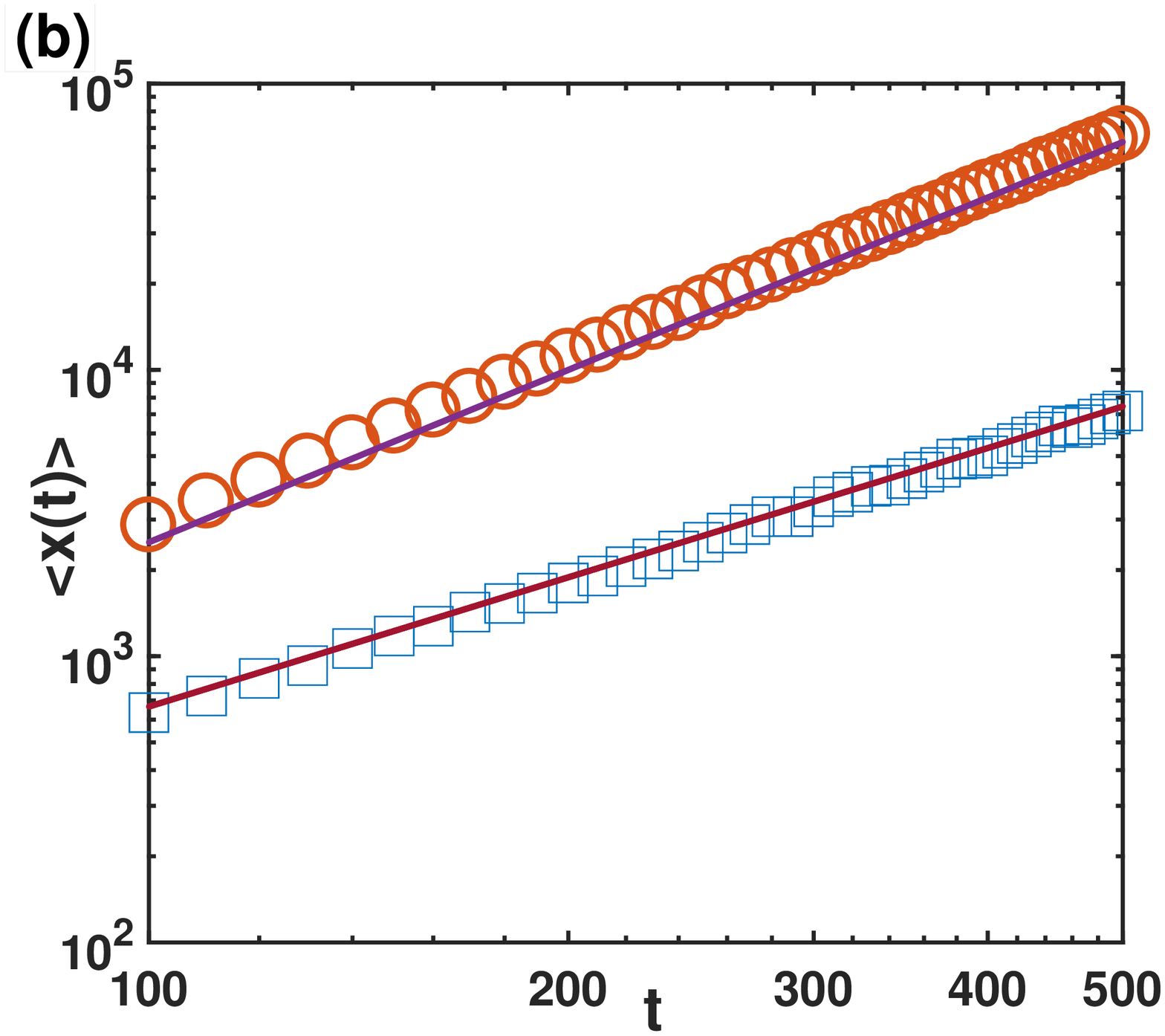}
  \caption{First moments of L\'{e}vy walk in a constant force field. 
 The running time density is, respectively, taken as exponential distribution and power-law distribution with $v_0=1$.
  We take $\lambda=1$ for exponential distribution (a) and $\alpha=0.5$ (squares) and $\alpha=1.5$ (circles), respectively, for power-law distribution with $\tau_0=1$ (b). These are obtained by averaging over $10^4$ realizations of L\'{e}vy walk in a constant force field. The solid lines are the theoretical results shown in \eqref{1.31} and \eqref{1.32}.}
  \label{CFthefirstmoment}
\end{figure}
\begin{figure}[htbp]
\centering
\includegraphics[scale=0.19]{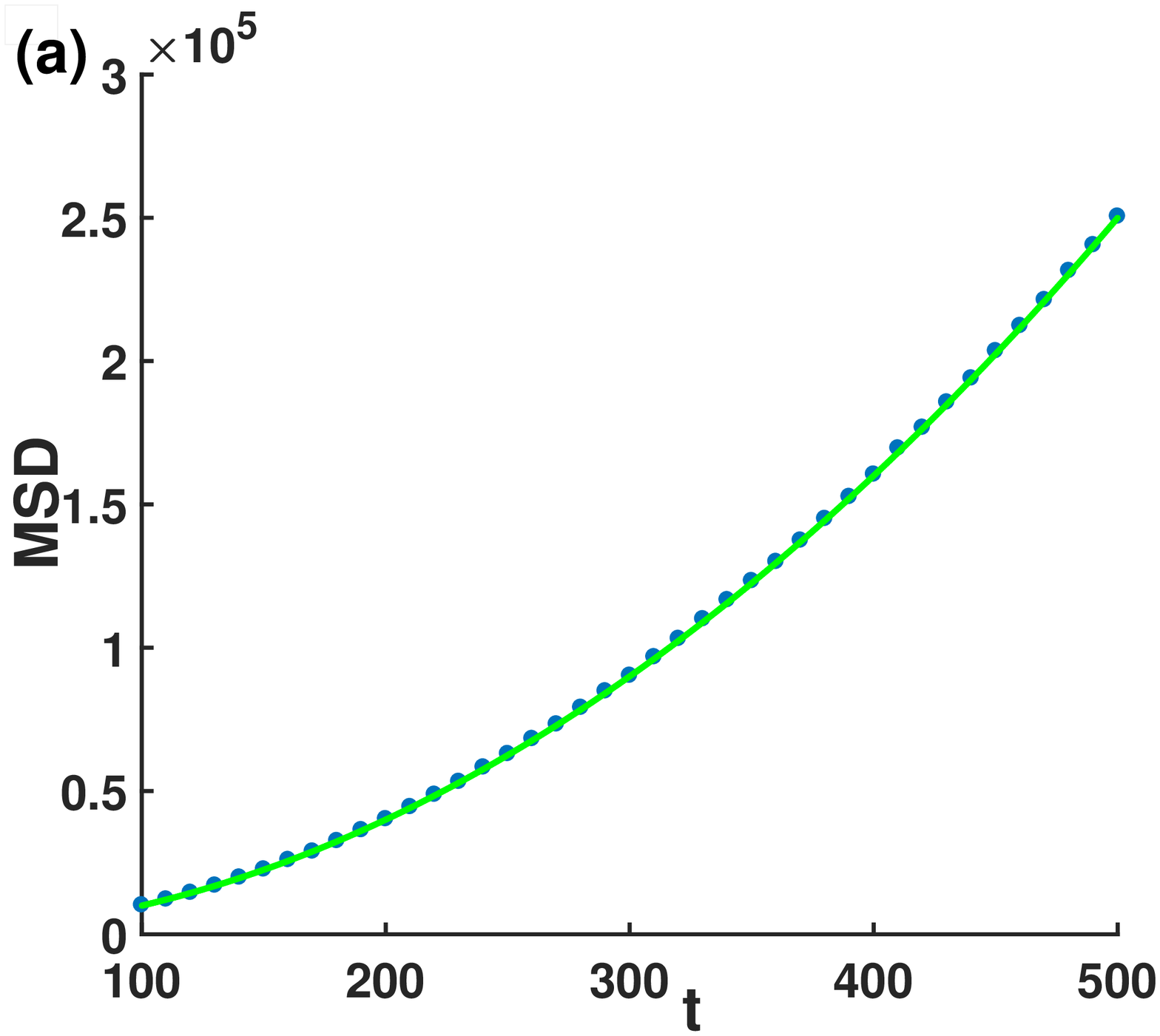}
\includegraphics[scale=0.19]{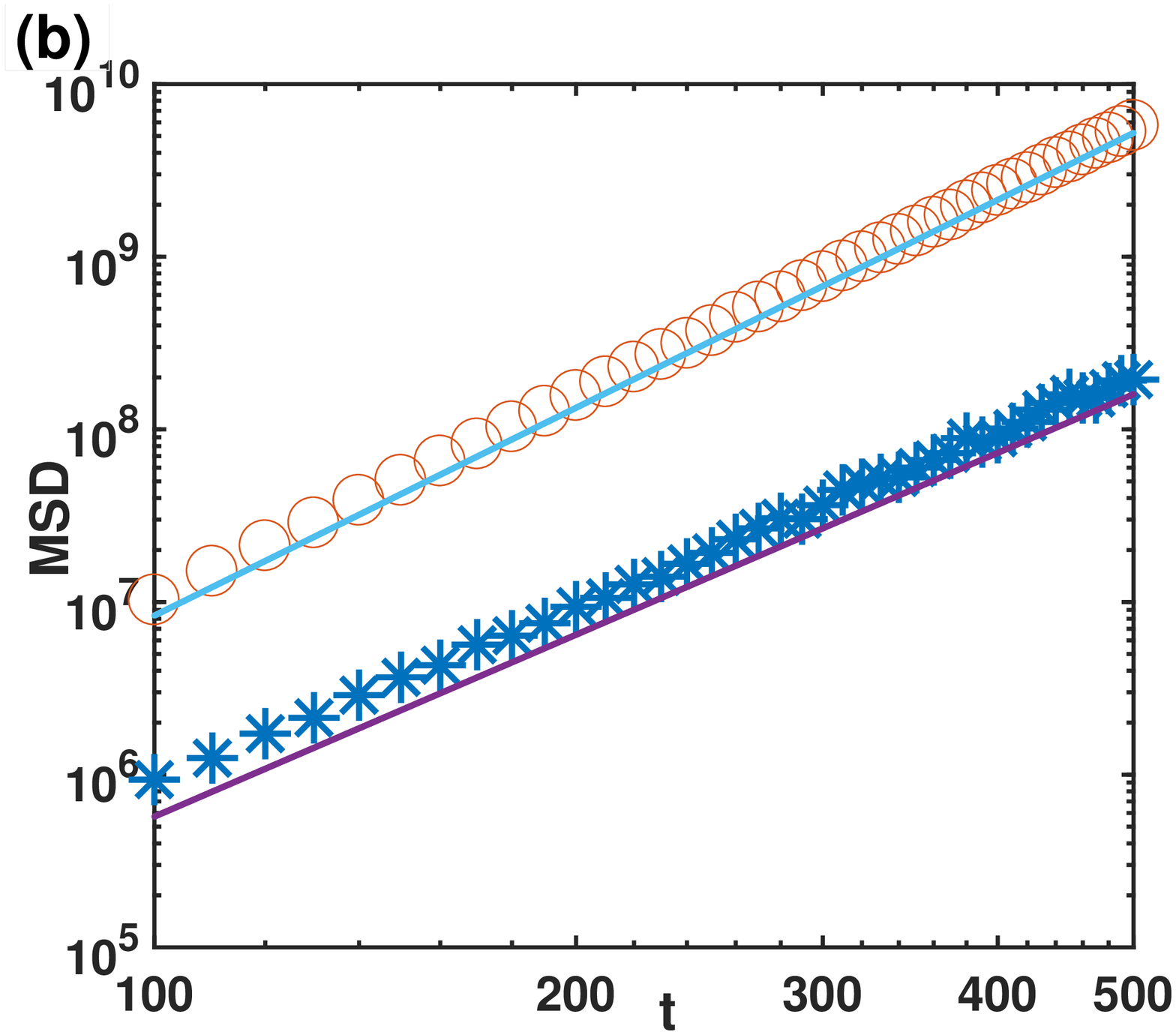}
\caption{MSD of L\'{e}vy walk in a constant force field with $v_0=1$ and $ x_0=0$. (a): $\phi(\tau)=\lambda e^{-\lambda \tau}$ with $\lambda=1$. (b): $\phi(\tau)=\frac{1}{\tau_0}\frac{\alpha}{(1+\tau/\tau_0)^{1+\alpha}}$ with $\tau_0=1$, and $\alpha=0.5$ (circles) and $\alpha=1.5$ (stars), respectively. The simulations are shown as dots with different marks by averaging over $10^4$ realizations and the solid lines are the corresponding theoretical results shown in \eqref{1.31} and Table \ref{tab1}. }
\label{CFMSD}
\end{figure}
The numerical simulations shown in Fig. \ref{CFthefirstmoment} and Fig. \ref{CFMSD}  verify the results. Here in this case, it is valuable to further consider the variance $\sigma(t)$ of the process, whose asymptotic form for sufficient long time $t$ is given by
\begin{equation}\label{var}
	\sigma(t)=\langle x^2(t)\rangle-\langle x(t)\rangle^2\sim \frac{2(a^2+\lambda^2 v_0^2)t}{\lambda^3}.
\end{equation}
Figure \ref{CFVarexpalpha1} verifies our result of variance. By taking $a=0$, it can be obtained from \eqref{var} that $\sigma(t)\sim 2 v_0^2 t/\lambda$, which recovers the MSD of classical symmetric L\'evy walk process  which has the constant value of velocity $v_0$ and the same running time distribution in one dimensional space with no external potential, in this sense the result shown in \eqref{var} indicates the constant force field can accelerate the L\'evy walk process. Besides it can be observed from \eqref{1.31} that the initial value of velocity $v_0$ for each step has no influence on the average displacement or the MSD when L\'evy walk particle moves in a constant force field after sufficient long time, and this conclusion can also be found in \eqref{1.32}.
%
%

Another representative kind of running time distribution is power-law, i.e.,
\begin{equation}\label{power_law}
	\phi(\tau)=\frac{1}{\tau_0}\frac{\alpha}{(1+\tau/\tau_0)^{1+\alpha}}
\end{equation}
with $\tau_0$ and $\alpha>0$, and the corresponding asymptotic form of Laplace transform is
\begin{equation}\label{waiting_time_appr}
  \hat{\phi}(s)\sim1-\frac{\tau_0}{\alpha-1}s-\tau_0^\alpha \Gamma(1-\alpha) s^\alpha+\frac{\tau_0^2 s^2}{(\alpha-1)(\alpha-2)}.
\end{equation}
By using the same techniques, the first two moments of the L\'evy walk in constant force field with power-law running time density can be calculated. The asymptotic behavior of average displacement for different range of $\alpha$ is
\begin{equation}\label{1.32}
  \langle x(t)\rangle=\begin{cases}
                        \frac{a(1-\alpha)}{2} t^2, & \mbox{if $0<\alpha<1$,}  \\
                        \frac{a (\alpha-1) \tau_0^{-1+\alpha}}{(3-\alpha)(2-\alpha)} t^{3-\alpha}, & \mbox{if $1<\alpha<2$}.
                      \end{cases}
\end{equation}
The asymptotic behaviors of MSDs are shown in Table \ref{tab1}, which are in accordance with the ones given in \cite{chen}. It should be noted that the results in \cite{chen} are obtained by using the Langvin pictures of L\'evy walk, while here we directly use the random walk theory to solve the problem.
\begin{table}
  \centering
  \begin{tabular}{|c|c|c|c|}
    \hline
    ~ & $0<\alpha<1$ & $1<\alpha<2$   \\
    \hline
    $a=0$ &  $(1-\alpha)v_0^2 t^2$ & $\frac{2 v_0^2 (\alpha-1)\tau_0^{-1+\alpha}}{(3-\alpha)(2-\alpha)}t^{3-\alpha}$  \\
    \hline
    $a\neq 0$ &  $\frac{a^2(1-\alpha)(3-2\alpha)}{12}t^4$ & $\frac{a^2 (\alpha-1) \tau_0^{-1+\alpha}}{(5-\alpha)(4-\alpha)} t^{5-\alpha}$ \\
    \hline
  \end{tabular}
  \caption{Asymptotic behavior of the MSD for L\'evy walk in constant force field. The running time density is power-law with the asymptotic behavior in Laplace space shown in \eqref{waiting_time_appr} and $\alpha$ is chosen in different regions.}\label{tab1}
\end{table}
For power-law running time distribution, it can be concluded from \eqref{1.32} and Table \ref{tab1} that the external constant force can accelerate L\'{e}vy particles  and $v_0$ does not affect the average movement and MSD for sufficient long time. The results are verified by Fig. \ref{CFthefirstmoment} and Fig. \ref{CFMSD}.


%

\begin{figure}[htbp]
  \centering
  \includegraphics[width=8cm]{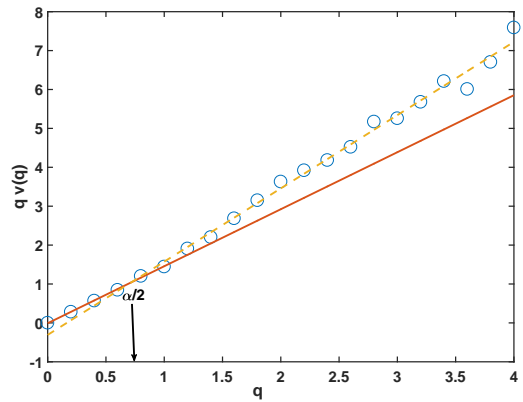}
  \caption{Power-law exponent w.r.t. $t$ of $\langle |x(t)|^q \rangle$ versus $q$ for $\alpha=3/2$. The piecewise linear property can be found in the simulations, specifically $q v(q)= q \alpha$ for $q < \alpha/2$ (solid line) and $ q v(q)= 2 q+1-\alpha $ for $q > \alpha/2$ (dashed line). The parameters are $v_0=a=1$ and $x_0=0$, and the simulation results are obtained by averaging over $10^4$ realizations.}
  \label{CFqmoments}
\end{figure}

In the final part of this section, we would like to discuss the fractional moments defined as \cite{MK2000,RDHB2014,SS2014}
\begin{equation}\label{qmoments}
  \langle |x(t)|^q \rangle=\int_{-\infty}^{+\infty} |x(t)|^q P(x,t)dx\sim M_q t^{q v(q)}.
\end{equation}
In fact, for Brownian motion $v(q)=1/2$ which is a constant. As referred to in \cite{CMMV1999}, the process with non-constant $v(q)$ is called a strongly anomalous diffusion. The symmetric sub-ballistic L\'evy walk with constant value of velocity moving in free one dimensional space belongs to this category because of the multiscaling property \cite{s14}. Next based on numerical simulations, we present the predictive results on the fractional moment and the property of strongly anomalous diffusion for the L\'evy walk process moving in a constant force field. 
By taking running time density function as power-law \eqref{power_law} with $1<\alpha<2$, the fractional moment can be obtained by numerical simulations shown in Fig. \ref{CFqmoments}, which is
\begin{equation*}
	\langle |x(t)|^q \rangle \sim \left\{
	\begin{split}
		&t^{\alpha q}, \quad \text{for}~q<\alpha/2,\\
		&t^{2q+1-\alpha}, \quad \text{for}~q>\alpha/2,
	\end{split}
	\right.
\end{equation*}
implying that L\'evy walk process moving in a constant force field is also a strongly anomalous diffusion. The turning point locates at $q=\alpha/2$, which is totally different from the case of free condition $q=\alpha$ \cite{s14}, besides the exponents for both regions of $q$ are also different. This is also one of the major influences made by the constant force field.

%
%

\section{L\'{e}vy Walk in the combined action of harmonic potential and constant force field }\label{sec 3}

L\'evy walk process in harmonic potential has been fully discussed in \cite{Pengbo1}. In this section we will consider L\'evy walk particle with mass $M$ moving in constant force field $F$ mixed with harmonic potential $V(x_t)=\frac{\gamma}{2} x_t^2$, where $\gamma>0$ is a constant.

\subsection{Introduction of the model}

Here we adopt the same notations as those in Sec. \ref{sec2_a}, and when L\'evy walk particle moves in mixed potential $V(x)=-Fx+\gamma x^2/2$, for each $j=1,\ldots,n$ with $t_j\leq t$ we have the dynamics
%
%
\begin{equation}\label{2.1}
\left\{
\begin{split}
	&M\frac{d^2 x_{t_{j}+\tau'}}{d {\tau'}^2}=F- \gamma x_{t_{j}+\tau'},~\text{if}~\tau' \in (0, \{t_{j+1}\wedge t\}-t_{j}],\\
	&\left.\frac{d}{d\tau'}x_{t_{j}+\tau'}\right|_{\tau'=0}=\pm v_0,
\end{split}
\right.
\end{equation}
where the initial velocity for each step is $\pm v_0$ and the probability of choosing either one of them is still $1/2$. For the initial velocity $\pm v_0$, the corresponding solution to \eqref{2.1} is
\begin{equation*}
	x_{t_{j}+\tau'}=\frac{F}{\gamma}+\left(x_{t_{j}}-\frac{F}{\gamma}\right)\cos(\omega\tau' )\pm \frac{v_0}{\omega}\sin(\omega\tau'),
\end{equation*}
where $\omega=\sqrt{\frac{\gamma}{M}}$. Similarly, the following equations for $q(x_t,t)$ and $P(x_t,t)$ can be given as
\begin{widetext}
\begin{align}
	q(x_t,t)=&\frac{1}{2}\int_{-\infty}^{\infty}\int_{0}^{t}q(x_{t-\tau},t-\tau)\phi(\tau)\delta\left(x_t-\frac{F}{\gamma}-\Big(x_{t-\tau}-\frac{F}{\gamma}\Big)\cos(\omega\tau)-\frac{v_0}{\omega}\sin(\omega\tau)\right)dx_{t-\tau}d\tau \nonumber \\
&+\frac{1}{2}\int_{-\infty}^{\infty}\int_{0}^{t}q(x_{t-\tau},t-\tau)\phi(\tau)\delta\left(x_t-\frac{F}{\gamma}-\Big(x_{t-\tau}-\frac{F}{\gamma}\Big)\cos(\omega\tau)+\frac{v_0}{\omega}\sin(\omega\tau)\right)dx_{t-\tau} d\tau +P_0(x)\delta(t),\label{2.2}\\
   P(x_t,t)=&\frac{1}{2}\int_{-\infty}^{\infty}\int_{0}^{t}q(x_{t-\tau},t-\tau)\Phi(\tau) \delta\left(x_t-\frac{F}{\gamma}-\Big(x_{t-\tau}-\frac{F}{\gamma}\Big)\cos(\omega\tau)-\frac{v_0}{\omega}\sin(\omega\tau)\right)dx_{t-\tau} d\tau \nonumber \\
     &+\frac{1}{2}\int_{-\infty}^{\infty}\int_{0}^{t}q(x_{t-\tau},t-\tau)\Phi(\tau)\delta\left(x_t-\frac{F}{\gamma}-\Big(x_{t-\tau}-\frac{F}{\gamma}\Big)\cos(\omega\tau)+\frac{v_0}{\omega}\sin(\omega\tau)\right)dx_{t-\tau}d\tau.\label{2.3}
\end{align}
Again we assume that $q$ and $P$ can be expressed by Hermite polynomials given in \eqref{1.10} and \eqref{1.11}, respectively; and when we choose $P_0(x)=\delta(x)$, the relations for $\widehat{T}_m(s)$ and $\widehat{R}_m(s)$ can be obtained as
\begin{align}
	 \sqrt{\pi} 2^m m! \widehat{T}_m (s)= &\frac{1}{2}\sum_{l=0}^{m}\sum_{j=0}^{\lfloor\frac{l}{2}\rfloor}\frac{\sqrt{\pi} m! 2^{l-2 j}}{(m-l)! j!}\widehat{T}_{l-2j}(s)\mathscr{L}_{\tau\rightarrow s}\Bigg\{\cos(\omega\tau)^{l-2 j}\big(-\sin(\omega\tau)\big)^{2 j} \phi(\tau)\nonumber\\
	 \times & \left[\left(-\frac{2 F}{\gamma}\cos(\omega\tau)+\frac{2 F}{\gamma}+\frac{2 v_0}{\omega}\sin(\omega\tau)\right)^{m-l}+\left(-\frac{2 F}{\gamma}\cos(\omega\tau)+\frac{2 F}{\gamma}-\frac{2 v_0}{\omega}\sin(\omega\tau)\right)^{m-l}\right]\Bigg\}+H_m(0),\label{2.11}\\
 \sqrt{\pi} 2^m m!\widehat{R}_m (s)= & \frac{1}{2}\sum_{l=0}^{m}\sum_{j=0}^{\lfloor\frac{l}{2}\rfloor}\frac{\sqrt{\pi} m! 2^{l-2 j}}{(m-l)! j!}\widehat{T}_{l-2j}(s)\mathscr{L}_{\tau\rightarrow s}\Bigg\{\cos(\omega\tau)^{l-2 j}\big(-\sin(\omega\tau)\big)^{2 j} \Phi(\tau)\nonumber\\
   & \left[\left(-\frac{2 F}{\gamma}\cos(\omega\tau)+\frac{2 F}{\gamma}+\frac{2 v_0}{\omega}\sin(\omega\tau)\right)^{m-l}+\left(-\frac{2 F}{\gamma}\cos(\omega\tau)+\frac{2 F}{\gamma}-\frac{2 v_0}{\omega}\sin(\omega\tau)\right)^{m-l}\right]\Bigg\}.\label{2.18}
\end{align}
\end{widetext}
The detailed derivations of \eqref{2.11} and \eqref{2.18} can be found in Appendix \ref{App_B}. Further the normalization of $P(x,t)$ can also be verified by $\widehat{T}_0(s)=\frac{1}{\sqrt{\pi}\left(1-\hat{\phi}(s)\right)}$ and $\widehat{R}_0(s)=\frac{1}{\sqrt{\pi} s}$ obtained from \eqref{2.11} and \eqref{2.18}, respectively. Next we will analyze some properties of L\'evy walk process in the mixed potential.

\subsection{Properties of L\'{e}vy walk process in harmonic potential combined with constant force field}

\subsubsection{Average displacement, MSD, and variance}

To begin with our discussion of this part, we still need to calculate $\widehat{T}_1(s)$ and $\widehat{R}_1(s)$ to obtain the average displacement $\langle x(t)\rangle$ following \eqref{firstmoment}; and in order to obtain the MSD $\langle x^2(t)\rangle$, the results of $\widehat{T}_2(s)$ and $\widehat{R}_2(s)$ are additionally necessary according to \eqref{msd}. From the relations \eqref{2.11} and \eqref{2.18}, by taking $m=1$ we have
%

\begin{equation}\label{2.14}
 \widehat{T}_1(s)=\frac{\widehat{T}_0(s)\Big[-\frac{2 F}{\gamma}\mathscr{L}_{\tau\rightarrow s}\big\{\cos(\omega\tau)\phi(\tau)\big\}+\frac{2 F}{\gamma}\hat{\phi}(s)\Big]}{2\Big[1-\mathscr{L}_{\tau\rightarrow s}\big\{\cos(\omega\tau)\phi(\tau)\big\}\Big]}
 \end{equation}
and
\begin{equation}\label{2.20}
\begin{split}
  \widehat{R}_1(s)&=\widehat{T}_0(s)\left[\frac{F}{\gamma}\hat{\Phi}(s)-\frac{F}{\gamma}\mathscr{L}_{\tau\rightarrow s}\big\{\cos(\omega\tau) \Phi(\tau)\big\}\right]\\
  &+\widehat{T}_1(s)\mathscr{L}_{\tau\rightarrow s}\big\{\cos(\omega\tau) \Phi(\tau)\big\};
  \end{split}
\end{equation}
taking $m=2$ leads to
\begin{equation}\label{2.16}
\begin{split}
   \widehat{T} &_2(s) =\frac{\sqrt{\pi}\widehat{T}_0(s)\mathscr{L}_{\tau\rightarrow s}\Big\{\phi(\tau)\big(\frac{4 F^2}{\gamma^2}\cos^2(\omega\tau)+\frac{4 F^2}{\gamma^2}\big)\Big\}}{8 \sqrt{\pi} \Big[1-\mathscr{L}_{\tau\rightarrow s}\big\{\cos^2(\omega\tau)\phi(\tau)\big\}\Big]}\\
   &+\frac{-\frac{8 F}{\gamma}\sqrt{\pi}\widehat{T}_1(s) \mathscr{L}_{\tau\rightarrow s}\Big\{\cos(\omega\tau)\phi(\tau)\big(\cos(\omega\tau)-1\big)\Big\}}{8\sqrt{\pi} \Big[1-\mathscr{L}_{\tau\rightarrow s}\big\{\cos^2(\omega\tau)\phi(\tau)\big\}\Big]} \\
      &+\frac{\sqrt{\pi}\widehat{T}_0(s) \mathscr{L}_{\tau\rightarrow s}\Big\{\phi(\tau)\big(-\frac{8 F^2}{\gamma^2}\cos(\omega\tau)\big)\Big\}-2}{8 \sqrt{\pi} \Big[1-\mathscr{L}_{\tau\rightarrow s}\big\{\cos^2(\omega\tau)\phi(\tau)\big\}\Big]}\\
       &+\frac{\sqrt{\pi}\widehat{T}_0(s)\mathscr{L}_{\tau\rightarrow s}\Big\{\phi(\tau)\big(\frac{4 v_0^2}{w^2}\sin^2(\omega\tau)-2\sin^2(\omega\tau)\big)\Big\}}{8 \sqrt{\pi} \Big[1-\mathscr{L}_{\tau\rightarrow s}\big\{\cos^2(\omega\tau)\phi(\tau)\big\}\Big]}
\end{split}
\end{equation}
and
\begin{equation}\label{2.21}
  \begin{split}
     &\widehat{R}_2(s) =\widehat{T}_0(s) \mathscr{L}_{\tau\rightarrow s}\left\{\Phi(\tau)\left(\frac{F^2}{2 \gamma^2}\cos^2(\omega\tau)+\frac{F^2}{2 \gamma^2}\right)\right\} \\
     &+\widehat{T}_0(s) \mathscr{L}_{\tau\rightarrow s}\left\{\Phi(\tau)\left(-\frac{F^2}{\gamma^2}\cos(\omega\tau)+\frac{ v_0^2}{2 w^2}\sin^2(\omega\tau)\right)\right\}\\
       & -\frac{F}{\gamma}\widehat{T}_1(s) \mathscr{L}_{\tau\rightarrow s}\left\{\cos^2(\omega\tau)\Phi(\tau)-\cos(\omega\tau)\Phi(\tau)\right\}\\
       & +\widehat{T}_2(s) \mathscr{L}_{\tau\rightarrow s}\big\{\cos^2(\omega\tau)\Phi(\tau)\big\}\\
       &+\widehat{T}_0(s) \mathscr{L}_{\tau\rightarrow s}\left\{ -\frac{1}{4}\sin^2(\omega\tau)\Phi(\tau)\right\}.
  \end{split}
\end{equation}

\begin{figure}[htbp]
\centering
\includegraphics[scale=0.19]{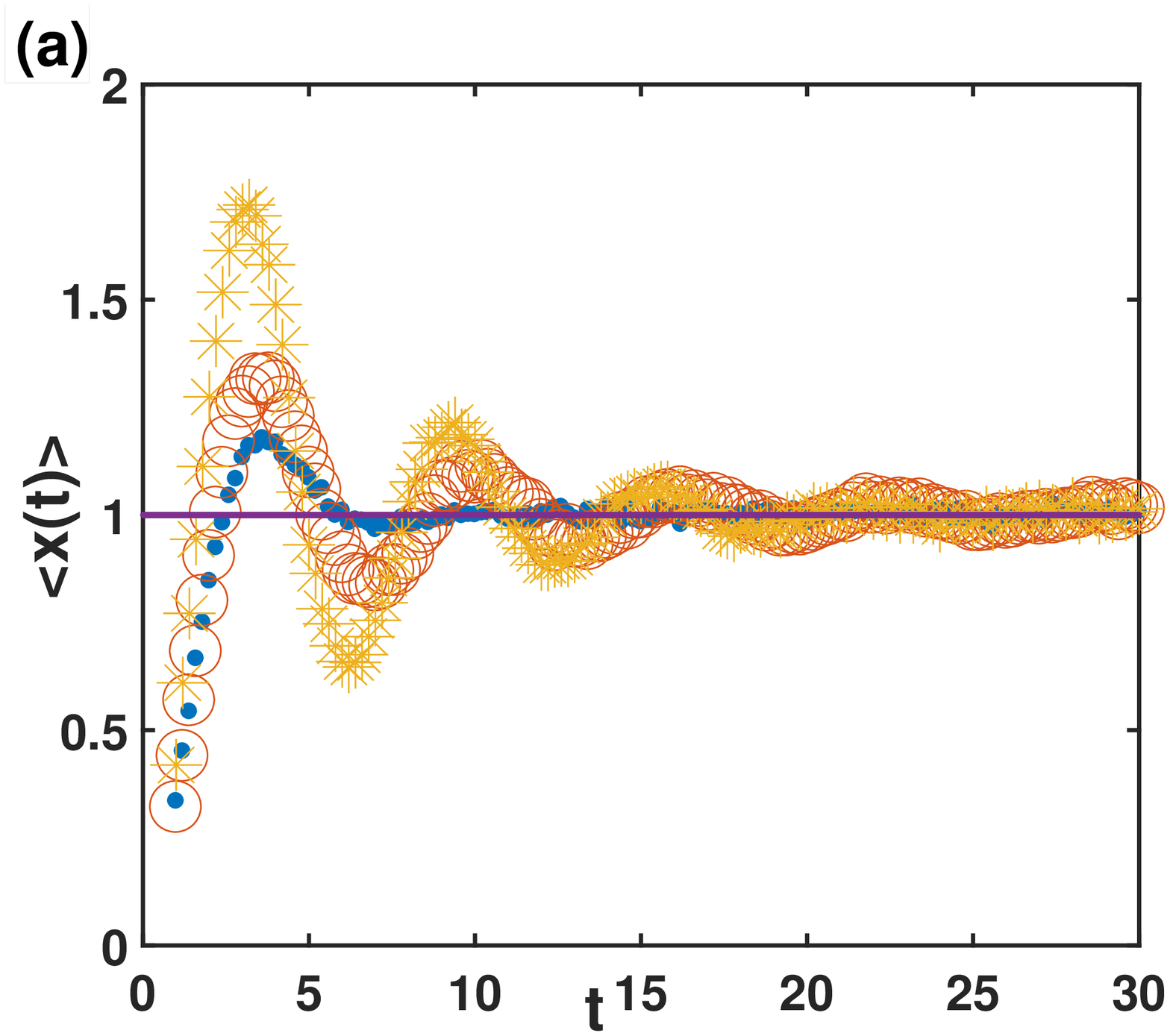}
\includegraphics[scale=0.19]{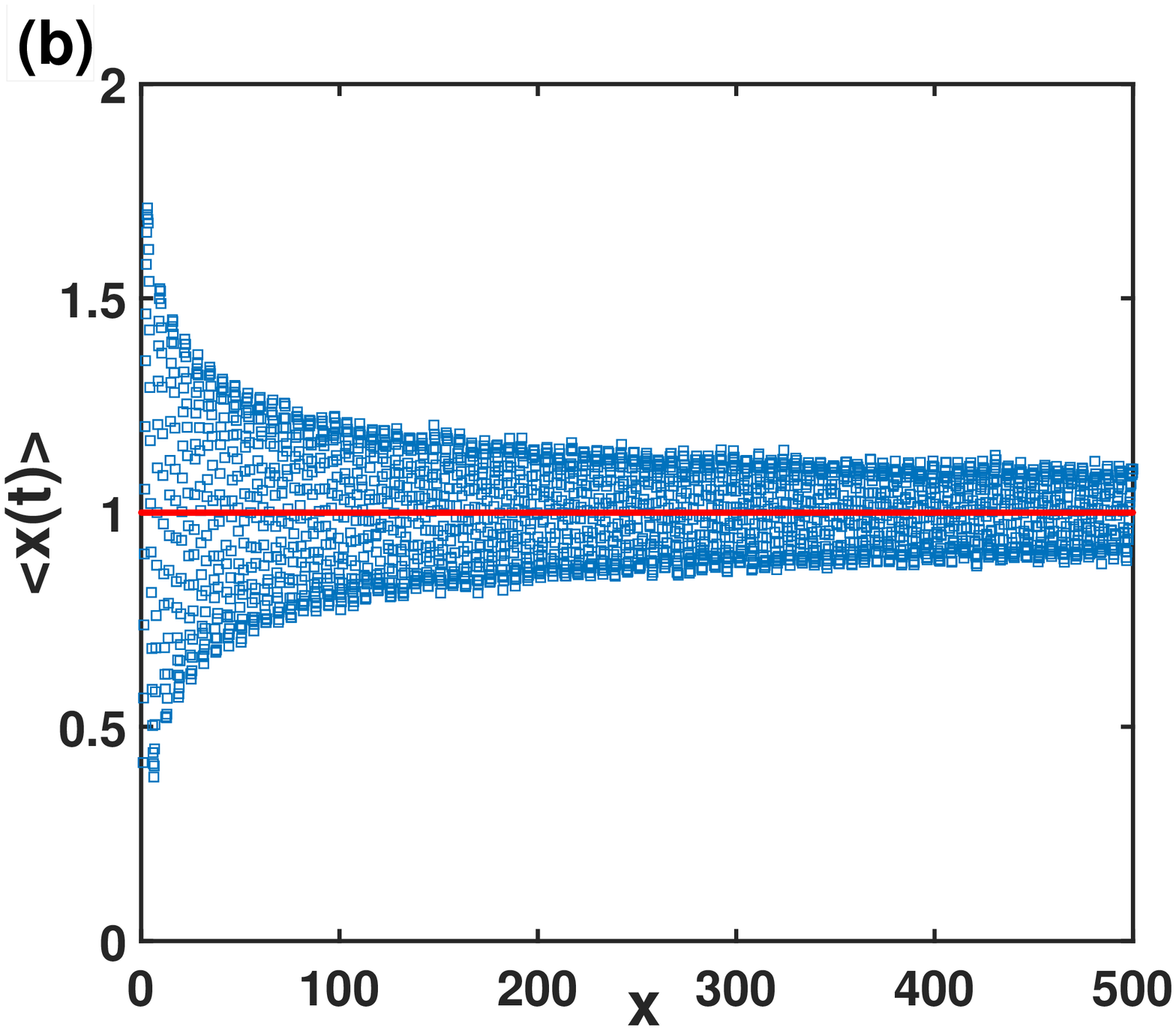}
\caption{Average displacement of L\'{e}vy walk in constant force field combined with harmonic potential with $v_0=1$. Figure (a) gives the simulation results sampling over $10^4$ realizations of $\phi(\tau)$ being exponential distribution with $\lambda=1$ (dots), power-law distribution with $\alpha=1.5$ (circles), and uniform distribution with period $T=2\pi$ (stars); for figure (b), $\phi(\tau)$ is taken as power-law distribution with $\alpha=0.5$. The solid lines are the theoretical result of $\langle x(t) \rangle$ shown in \eqref{avg_F+har}.}
\label{CFHPfirstmoment}
\end{figure}

\begin{figure}[htbp]
\centering
\includegraphics[width=4cm]{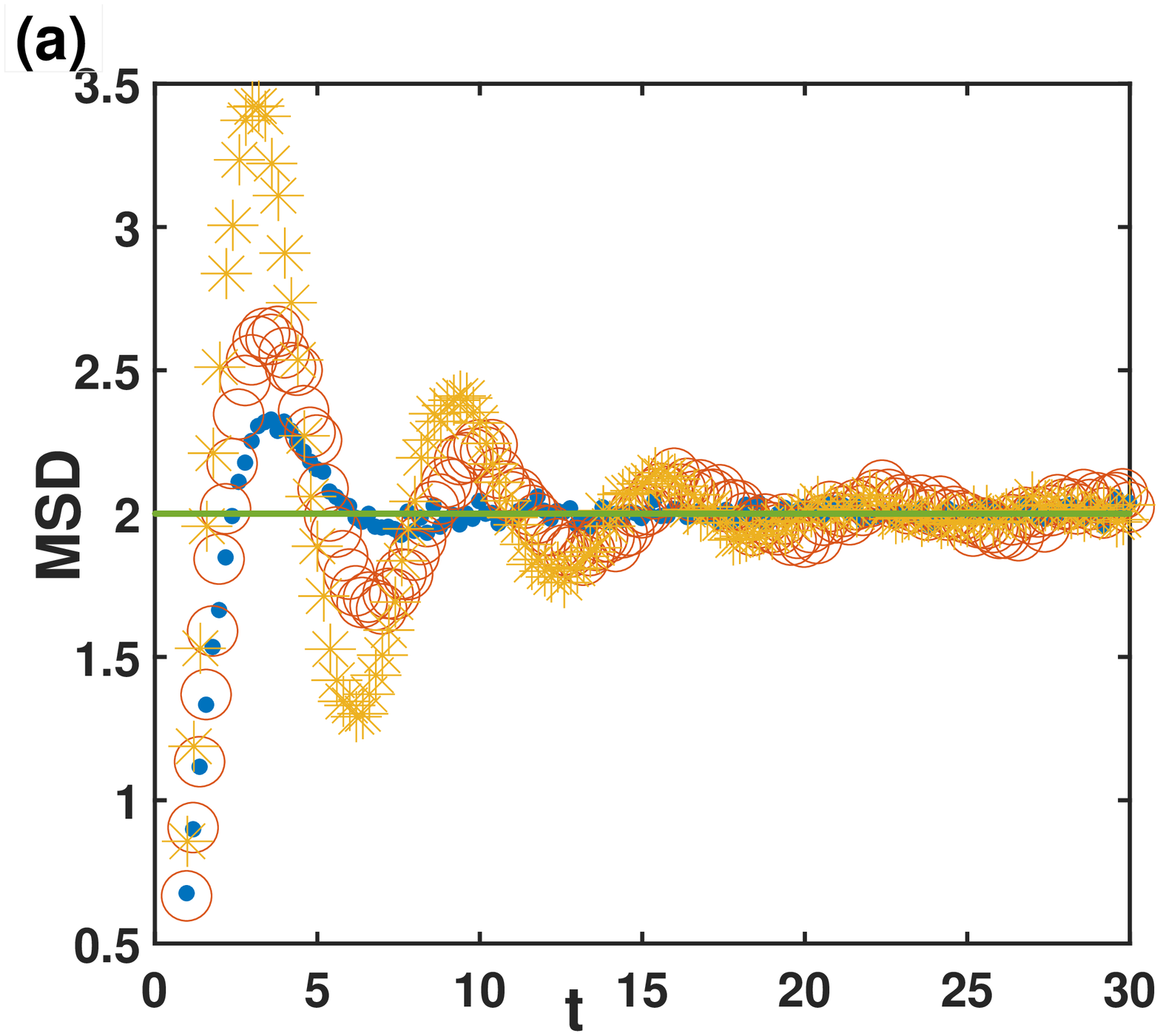}
\includegraphics[width=4cm]{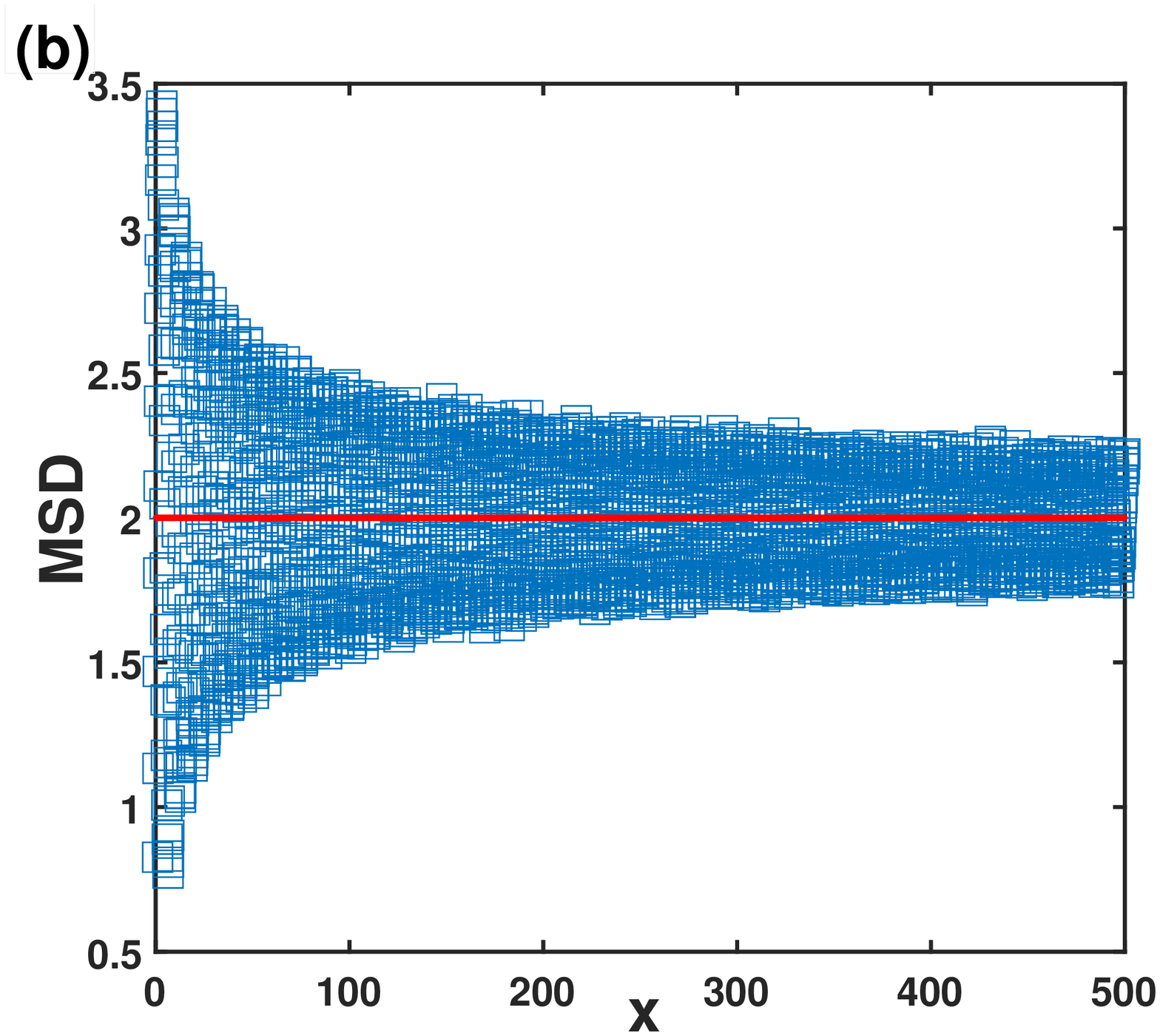}
\caption{MSD of L\'{e}vy walk in constant force field combined with harmonic potential with $v_0=1$. The parameters and the symbols are the same as the ones of Fig. \ref{CFHPfirstmoment}, and the solid lines represent the corresponding theoretical result of $\langle x^2(t)\rangle$ in \eqref{msd_F+har}. From the simulation results, one can see that the MSDs will eventually converge to the theoretical result when time $t$ is long enough.}
\label{CFHPMSD}
\end{figure}

For the first case, we consider exponentially distributed running time, i.e., $\phi(\tau)=\lambda e^{-\lambda\tau}$. Substituting the corresponding Laplace transform of $\phi(\tau)$ into \eqref{2.14} and \eqref{2.20}, then the asymptotic form $\langle x(t)\rangle$ for sufficiently long time $t$ in Laplace space has the form $\langle \hat{x}(s)\rangle \sim\frac{F}{\gamma s}$,
which after inverse Laplace transform leads to
\begin{equation}\label{avg_F+har}
	\langle x(t)\rangle \sim\frac{F}{\gamma}.
\end{equation}
From \eqref{2.16} and \eqref{2.21}, there exists
\begin{equation*}
	\langle\hat{x}^2(s)\rangle \sim \frac{1}{s}
	\left(\frac{v_0^2}{\omega^2}+\frac{F^2}{\gamma^2}\right);
\end{equation*}
and the corresponding inverse Laplace transform yields
\begin{equation}\label{msd_F+har}
	\langle x^2(t)\rangle \sim \frac{v_0^2}{\omega^2}+\frac{F^2}{\gamma^2}.
\end{equation}
Therefore the corresponding variance is
\begin{equation}\label{var_F+har}
	\sigma(t)\sim\frac{v_0^2}{\omega^2}.
\end{equation}
In fact, we can further choose $\phi(\tau)$ to be the power-law distribution and the uniform distribution on the interval $[0,T]$, i.e., $\phi(\tau)=\frac{1}{T}\mathbf{1}_{[0,T]}(\tau)$ with $T$ being the period of the process $T=2\pi/\omega$ and $\mathbf{1}_{[0,T]}(\tau)$ being the indicator function. Then the corresponding average movements and MSDs have the same asymptotic forms as shown in \eqref{avg_F+har} and \eqref{msd_F+har}, so that the asymptotic behaviors of variances are also the same as \eqref{var_F+har}. These results are verified by Fig. \ref{CFHPfirstmoment} and Fig. \ref{CFHPMSD}. From the result of MSD in \eqref{msd_F+har}, we conclude that L\'evy walk particles in constant force field combined with Harmonic potential are always localized, and this conclusion also reflects that in some sense harmonic potential is stronger than constant force field. Besides, it can also be concluded that the variance does not depend on the constant force $F$, and $\sigma(t)$ is the same as the MSD of L\'evy walk in potential $V(x)=\frac{\gamma}{2}x^2$ in \cite{Pengbo1}.


%

\subsubsection{Stationary distribution and kurtosis}

\begin{figure}[htbp]
\centering
\includegraphics[scale=0.25]{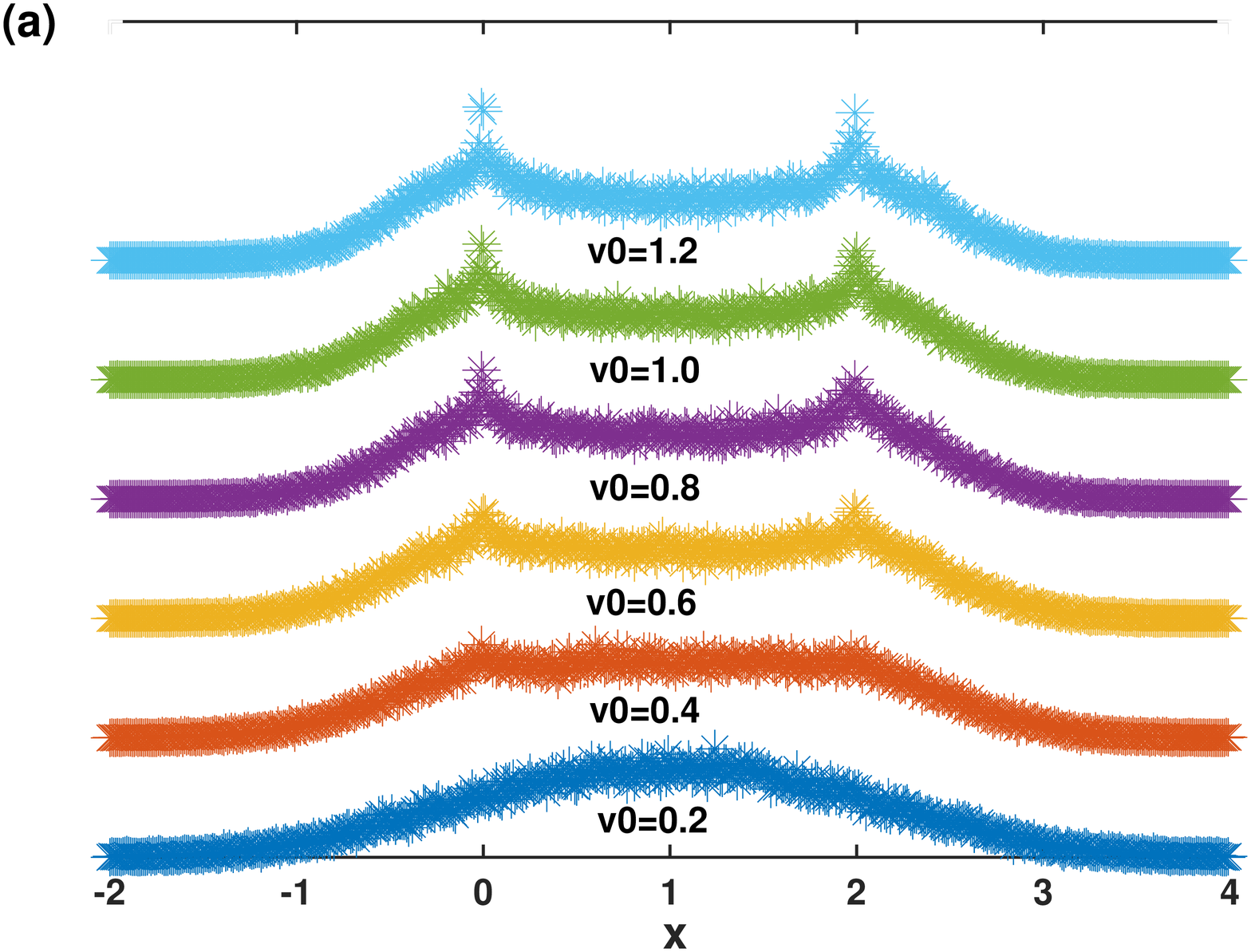}
\includegraphics[scale=0.25]{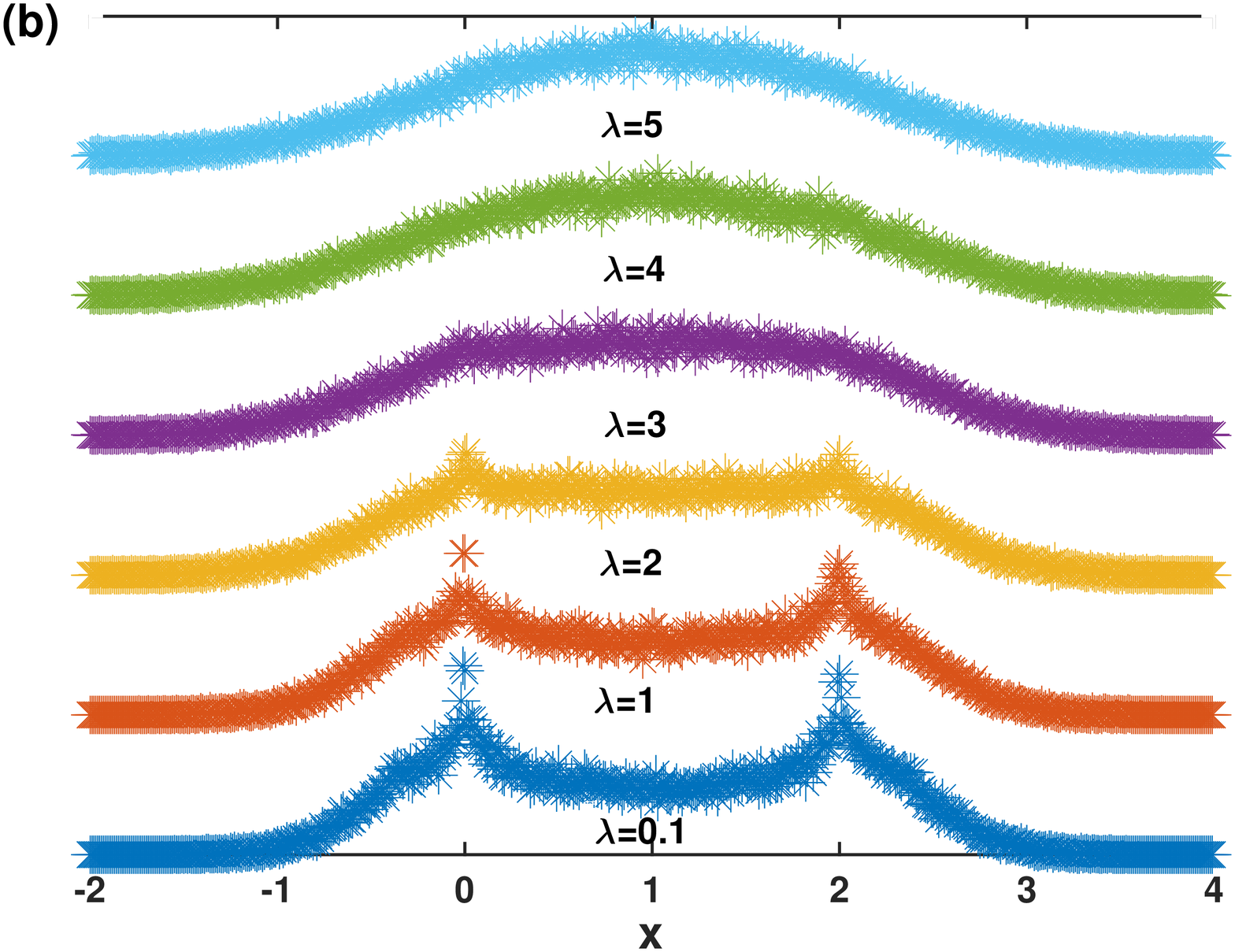}
\caption{Stationary PDFs of L\'evy walk with $\phi(\tau)=\lambda e^{-\lambda\tau}$ in constant force field combined with harmonic potential from numerical simulations. Figures (a) and (b) are, respectively, with fixed $\lambda=1$ and fixed $v_0=1$. The other parameters are $v_0=\omega=F=\gamma=1$.}
\label{CFHPPSTexp}
\end{figure}

\begin{figure}[htbp]
\centering
\includegraphics[width=4.2cm]{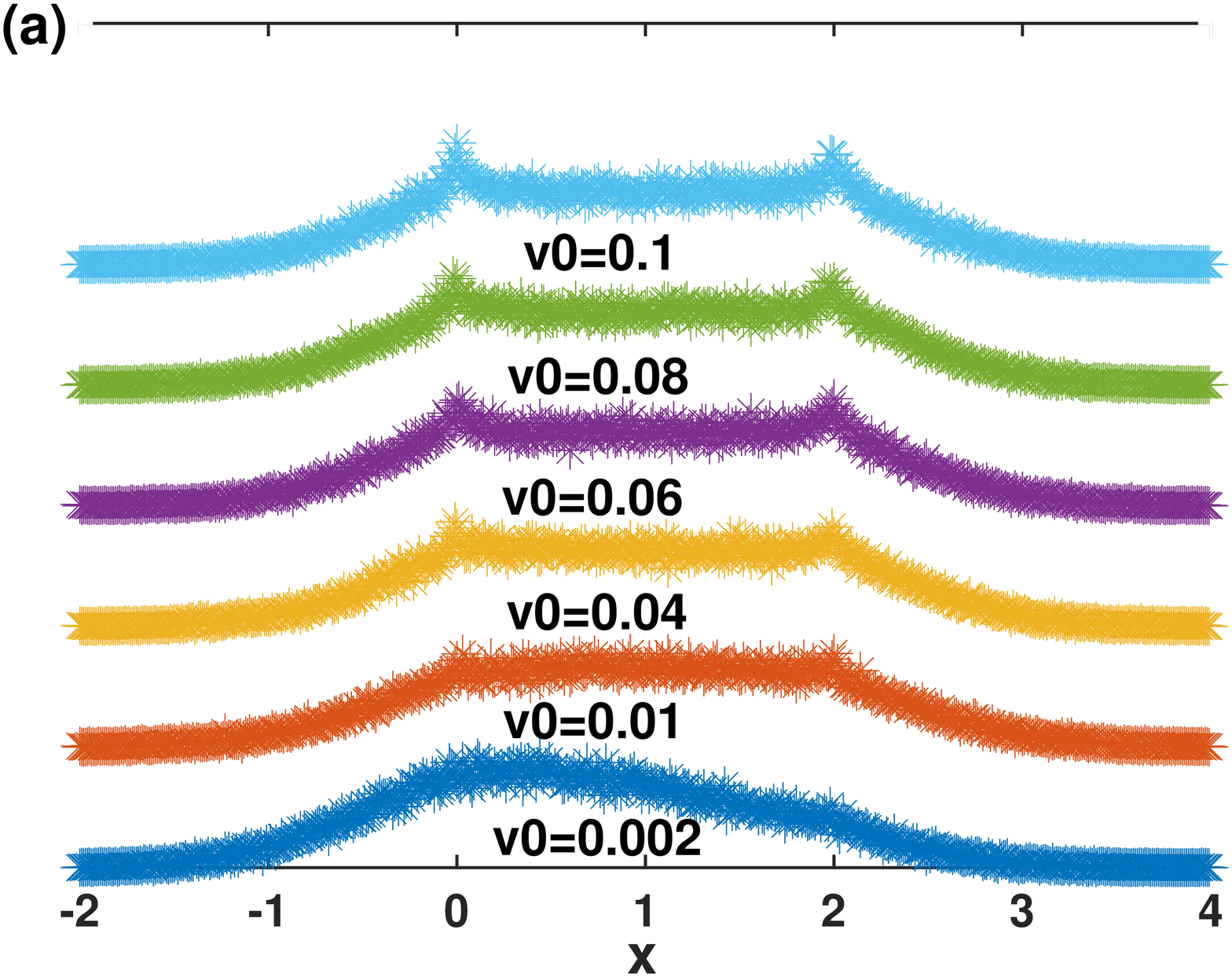}
\includegraphics[width=4.2cm]{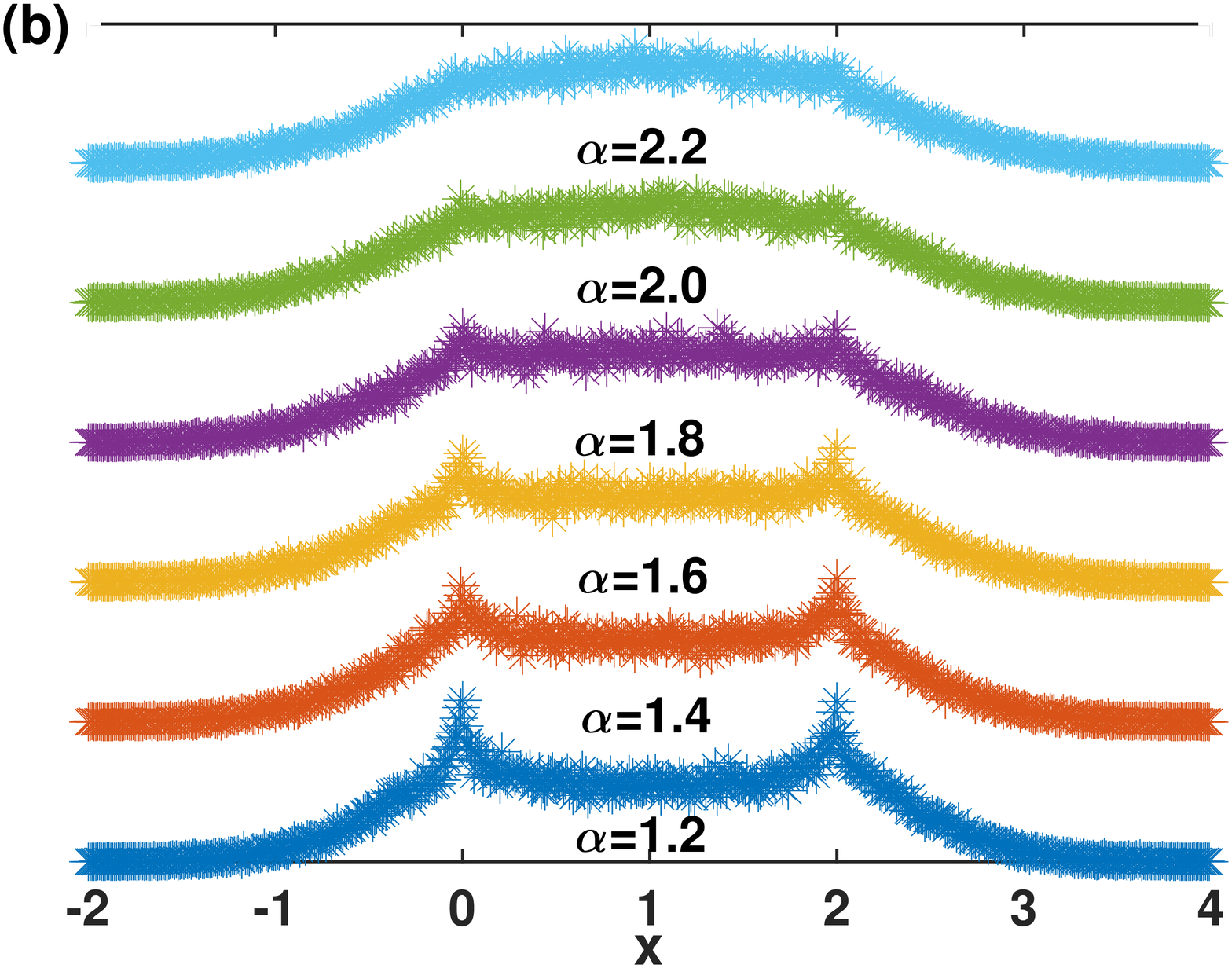}
\includegraphics[width=4.2cm]{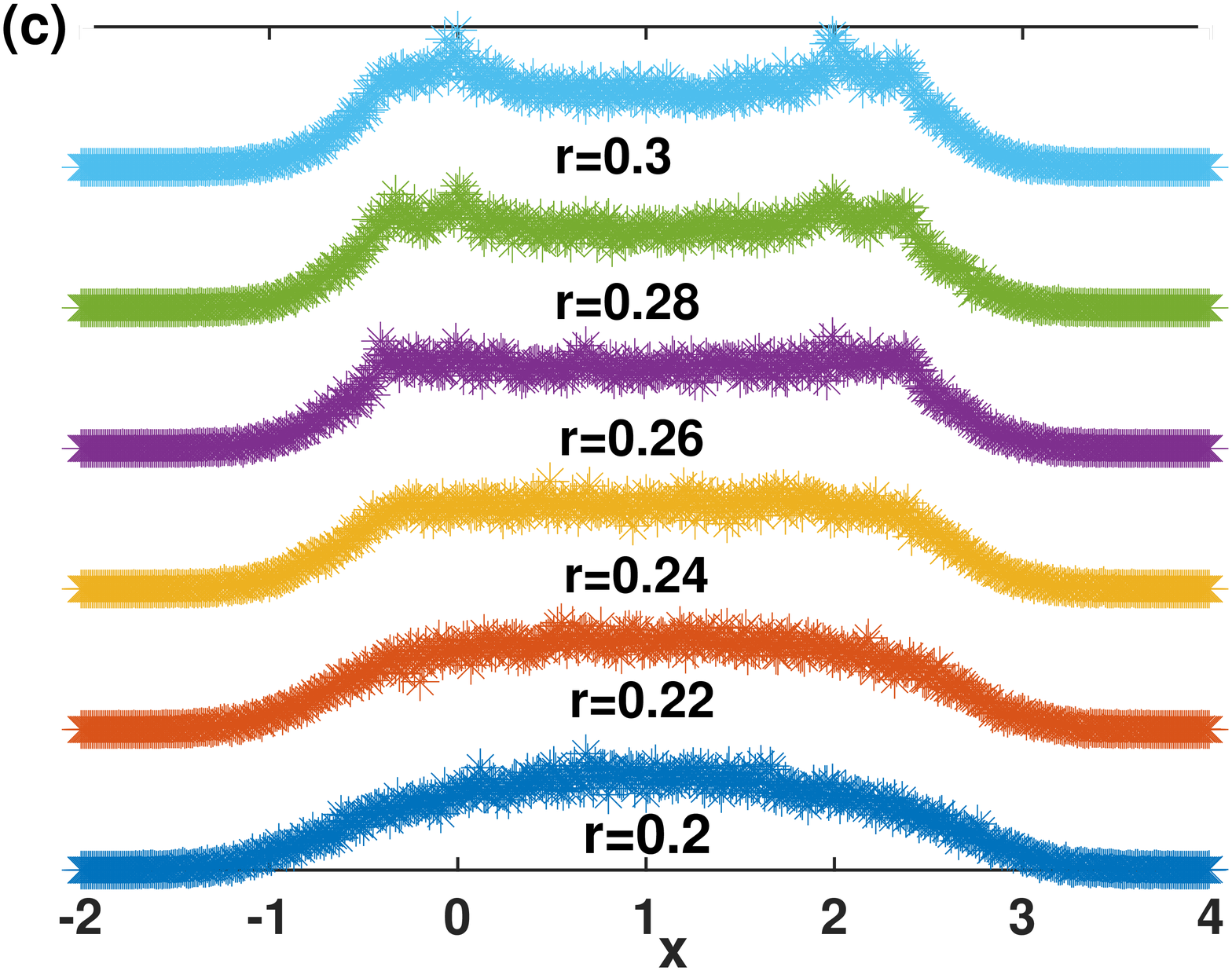}
\includegraphics[width=4.2cm]{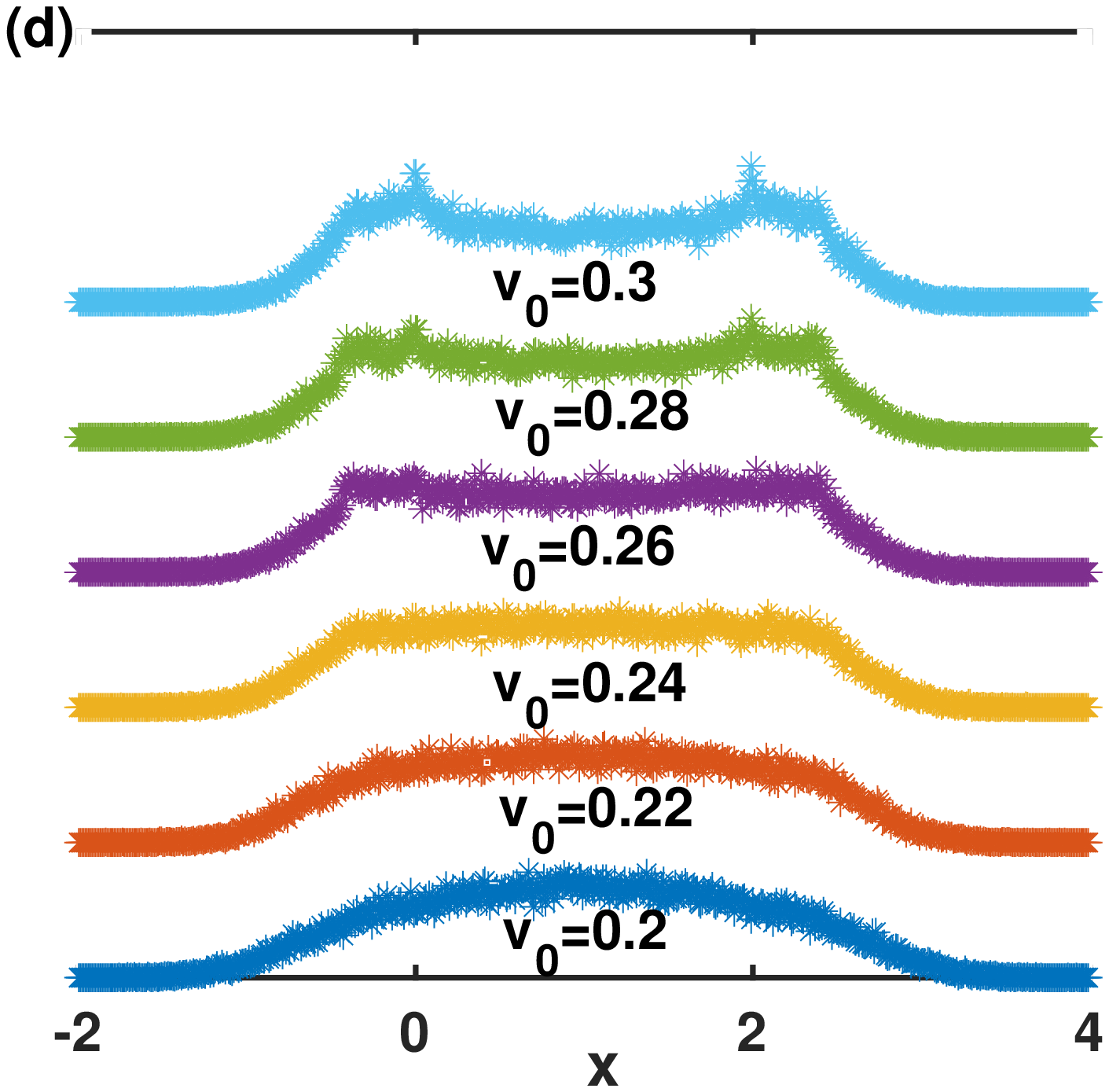}
\caption{ Stationary PDFs of L\'evy walk for power-law and uniformly distributed running times.
For all figures, $F=\gamma=1$ and $v_0=\omega$. For (a) and (b), we take $\phi(\tau)=\frac{\alpha}{(1+\tau)^{1+\alpha}}$ with $\alpha=1.5$ in (a) and $v_0=0.1$ in (b), respectively. For (c) and (d), the running time follows $\phi(\tau)=\frac{\omega}{2 \pi r}\mathbf{1}_{[0,\frac{2 \pi r}{\omega}]}(\tau)$ with $v_0=\omega=1$ in (c) and the relation $v_0=\omega=r$ in (d), respectively.}
\label{Pst_2}
\end{figure}

From the discussions in previous parts, the conclusion of localization indicates that further study on stationary distribution makes sense. The phenomena of monomodal-to-bimodal crossovers can be observed from Fig. \ref{CFHPPSTexp} and Fig. \ref{Pst_2}. Besides the peaks of bimodal distribution locate at $x\approx \frac{F}{\gamma}\pm \frac{v_0}{\omega}$, while for pure harmonic potential $V(x)=\gamma x^2/2$ the peaks locate at $x\approx \pm v_0/\omega$ for binomal state \cite{Pengbo1}. It is also a major difference from L\'evy flight process in an external potential, whose stationary distribution shows a bimonal state only for the potential steeper than harmonic one  \cite{CKGM2003,CGKM2005}.


For L\'evy walk with $\phi(\tau)=\lambda e^{-\lambda\tau}$ moving in constant force field combined with harmonica potential, the bimodal or monomodal state of stationary distribution $P^{st}(x)$ depends on parameter $\lambda$ and value of initial velocity $v_0$ for each step; to be more specific, as shown in Fig. \ref{CFHPPSTexp} the larger $v_0$ or smaller $\lambda$ is the bimodal state will emerge, for smaller $v_0$ or larger $\lambda$ the monomodal state can be observed. For $\phi(\tau)$ being a power-law density, the bimonal stationary distribution can also be observed when $v_0$ is larger or $\alpha$ is smaller, which can be found from numerical simulations shown in Fig. \ref{Pst_2}(a) and (b). Besides it can be concluded from Fig. \ref{Pst_2}(c) and (d) that for $\phi(\tau)=\frac{\omega}{2 \pi r}\mathbf{1}_{[0,\frac{2 \pi r}{\omega}]}(\tau)$ the bimonal state will appear when $r$ or $v_0$ becomes larger. Comparing with the results of pure harmonic potential in \cite{Pengbo1}, the combined potential with constant force field will bring the corresponding stationary distribution a translation of $F/\gamma$ and a skewness.

\begin{figure}[htbp]
\centering
\includegraphics[width=4cm]{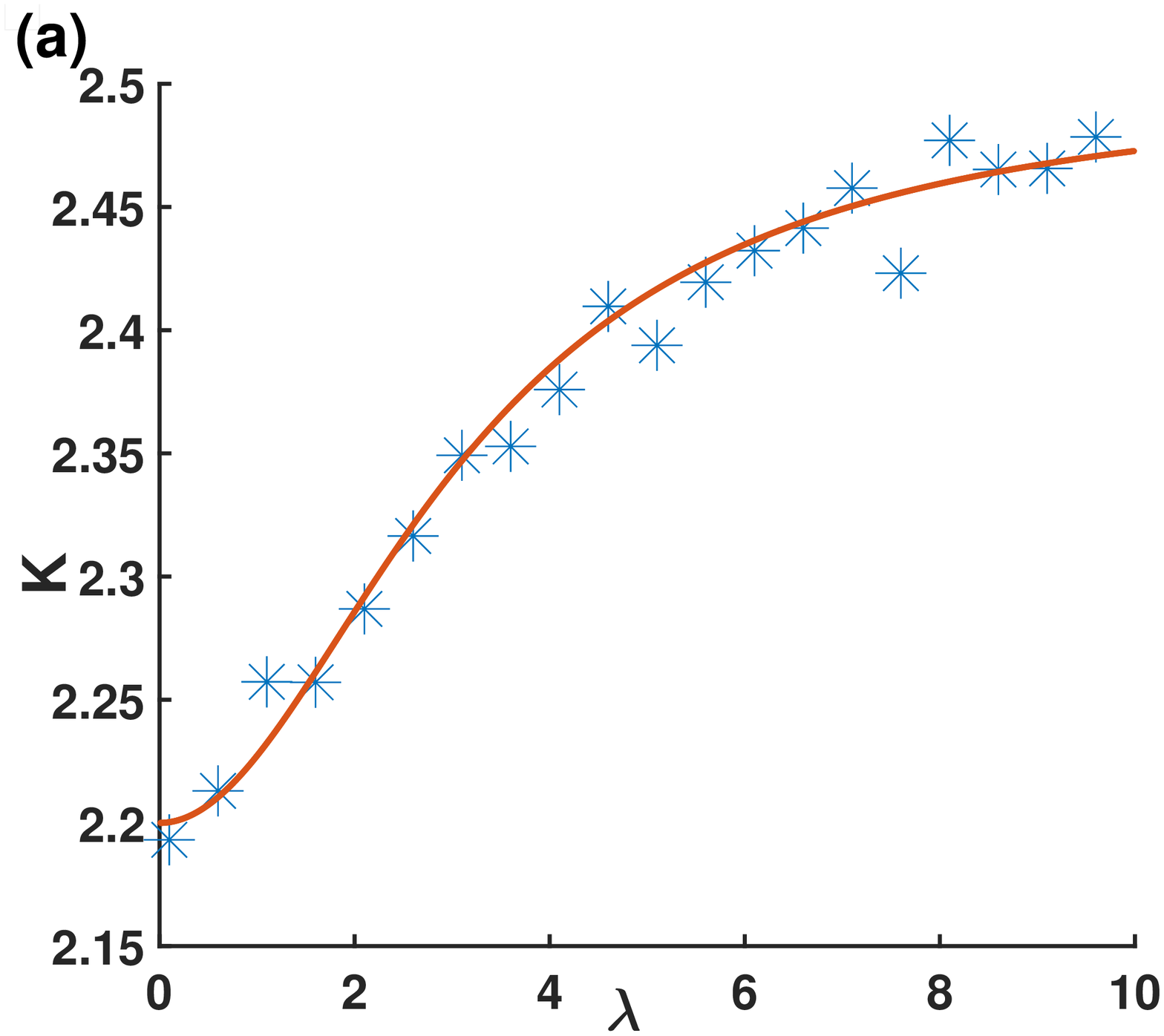}
\includegraphics[width=4cm]{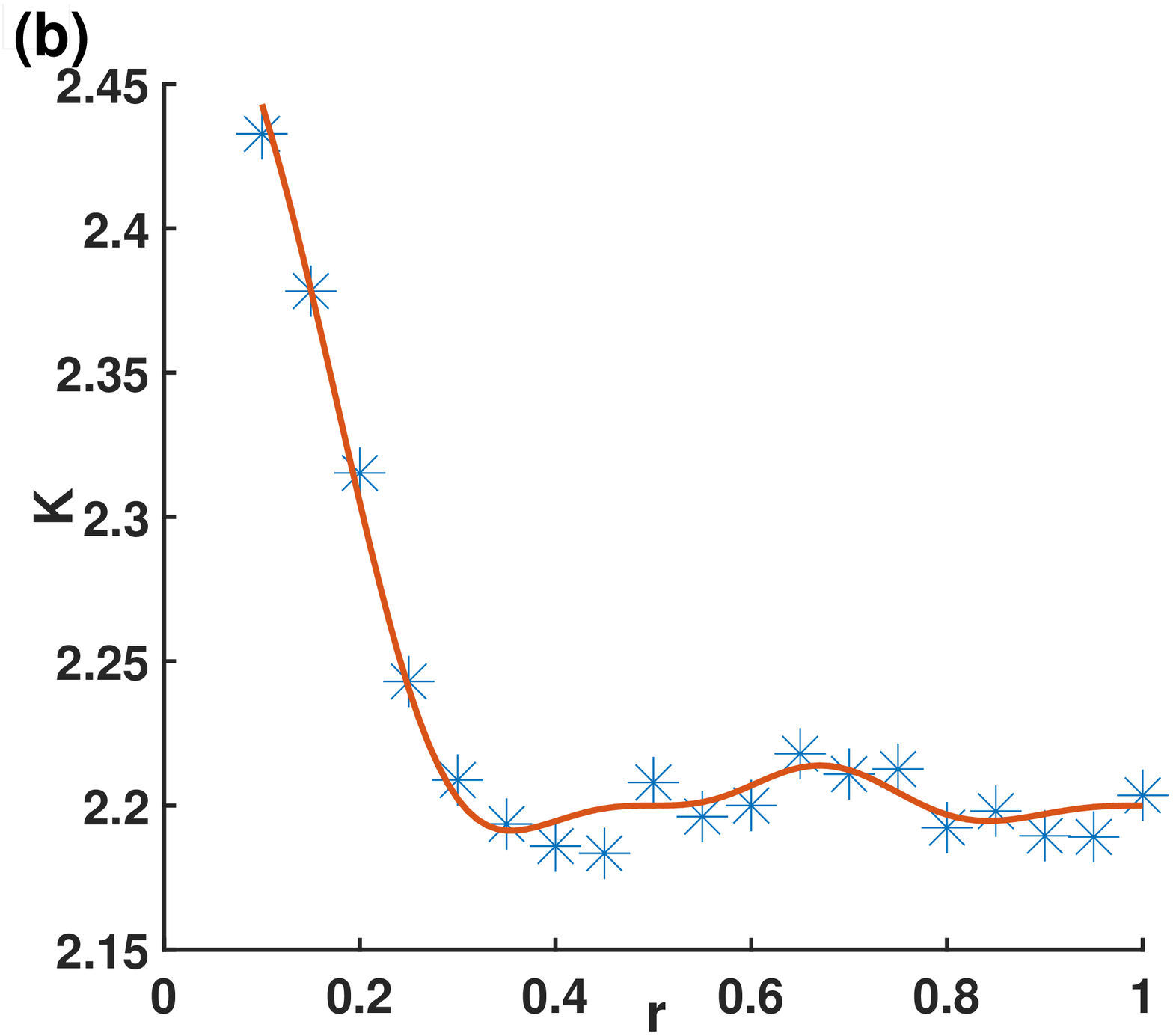}
\includegraphics[width=4cm]{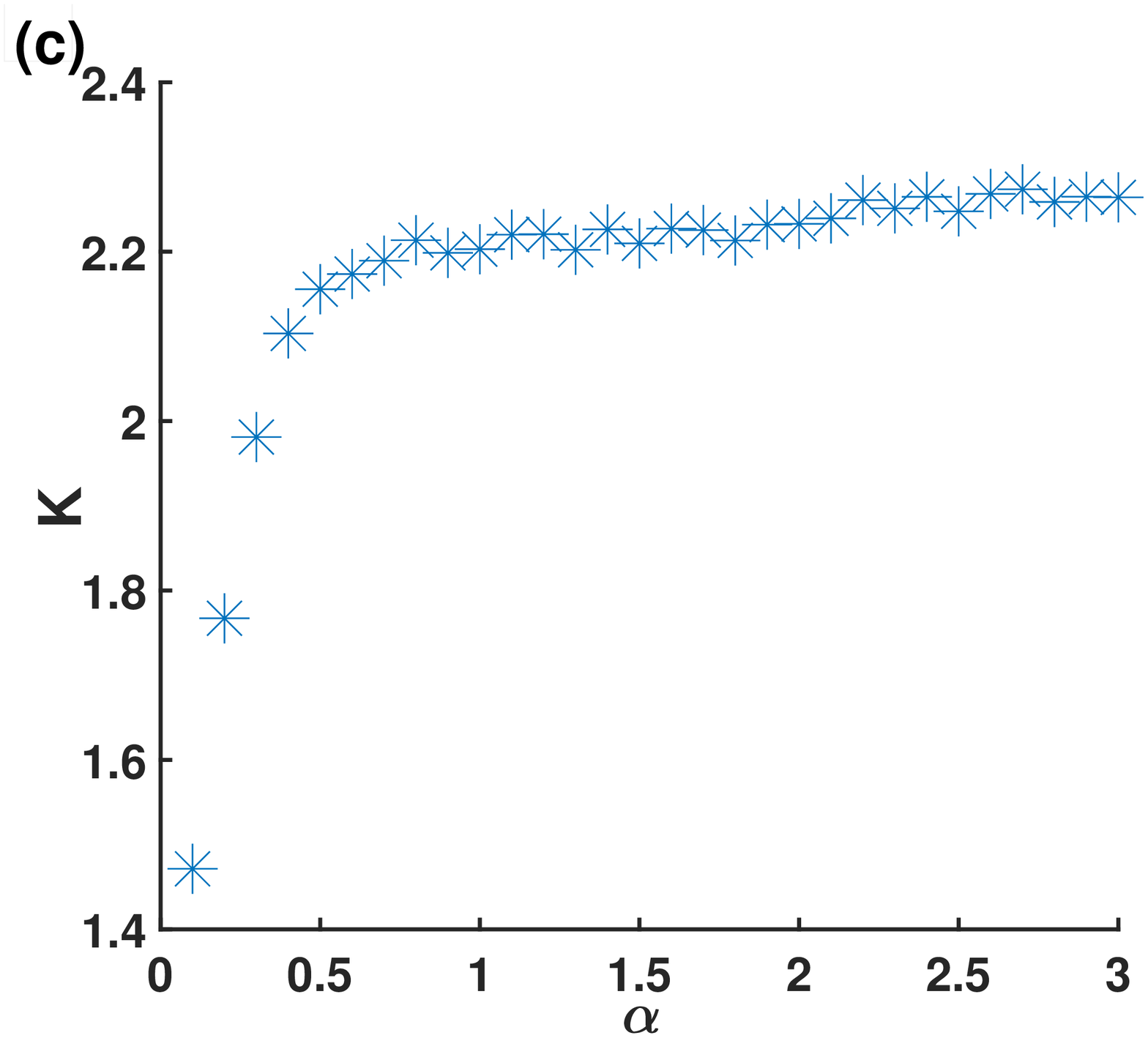}
\caption{Simulation results of the kurtosis $K$ for three different $\phi(\tau)$ sampling over $5\times10^{4}$ realizations at time $t=10^{4}$. The parameters are $v_0=\omega=\gamma=F=1$. For (a),  $\phi(\tau)=\lambda e^{-\lambda\tau}$, and the figure is with $K$ versus $\lambda$; for (b),  $\phi(\tau)=\frac{\omega}{2 \pi r}\mathbf{1}_{[0,\frac{2 \pi r}{\omega}]}(\tau)$, and the figure is with $K$ versus $r$; for (c), $\phi(\tau)=\frac{1}{\tau_0}\frac{\alpha}{(1+\tau/\tau_0)^{1+\alpha}}$, and the figure is with $K$ versus $\alpha$. The solid lines are for the numerical inverse Laplace transform of the exact form of $K$ in Laplace space for exponential and uniform $\phi(\tau)$.}
\label{K}
\end{figure}

Another important quantity which characterizes the tails of the stationary PDF is kurtosis defined as
\begin{equation*}
	K=\langle x^4(t)\rangle/\langle x^2(t)\rangle^2.
\end{equation*}
The fourth moment can be obtained through  \eqref{1.27} and \eqref{mmt}, which is
\begin{equation*}
	\langle\hat{x}^4(s)\rangle=3/4 \sqrt{\pi}\widehat{R}_0(s)+6\sqrt{\pi}\widehat{R}_2(s)+24 \sqrt{\pi}\widehat{R}_4(s).
\end{equation*}
 When $\phi(\tau)=\lambda e^{-\lambda\tau}$, \eqref{2.11} and \eqref{2.18} lead to
 \begin{equation}\label{exp_4mmt_F+har}
 	\langle\hat{x}^4(s)\rangle\sim \frac{1}{s}\left(\frac{F^4}{\gamma^4}+\frac{6 F^2 v_0^2}{\gamma^2 \omega^2}+\frac{3 v_0^4 (\lambda^2+6 \omega^2) }{\omega^4 (\lambda^2+10 \omega^2)}\right).
 \end{equation}
Applying inverse Laplace transform on \eqref{exp_4mmt_F+har} and combining with \eqref{msd_F+har}, we have
\begin{equation}\label{3.1}
\begin{split}
  K\sim  & \frac{3 \gamma^4 v_0^4(\lambda^2+6 \omega^2)}{(\lambda^2+10 \omega^2)(\gamma^2 v_0^2+F^2 \omega^2)^2}\\
     & +\frac{6 F^2 \gamma^2 v_0^2 \omega^2 +F^4 \omega^4}{(\gamma^2 v_0^2+F^2 \omega^2)^2},
\end{split}
\end{equation}
after sufficiently long time. The specific form of $K$ is verified by Fig. \ref{K} through numerical inverse Laplace transform; besides by observing the value of $K$ we can also conclude that the additional external constant force $F$ cannot change the platykurtic state for small $\lambda$, and from \eqref{3.1} we have the limit
\begin{equation*}
	\begin{split}
		\lim_{\lambda\to\infty}K&\simeq\frac{3\gamma^4 v_0^4+6F^2\gamma^2v_0^2\omega^2+F^4\omega^4}{(\gamma^2 v_0^2+F^2\omega^2)^2}\\
		&=3-\frac{2F^4\omega^4}{(\gamma^2 v_0^2+F^2\omega^2)^2}<3,
	\end{split}
\end{equation*}
which indicates the stationary distribution is always platykurtic if $F\neq 0$ and this conclusion is a major difference from pure harmonic case as shown in \cite{Pengbo1}. Therefore we will not expect a Gaussian distribution when $F\neq 0$ for large $\lambda$. Besides for the uniform distribution $\phi(\tau)=\frac{\omega}{2 \pi r}\mathbf{1}_{[0,\frac{2 \pi r}{\omega}]}(\tau)$, the asymptotic behavior of $K$ is

\begin{equation}\label{3.111}
  K\sim\frac{3 \gamma^4 v_0^4+6 F^2 \gamma^2 v_0^2 \omega^2 +F^4\omega^4}{(\gamma^2 v_0^2+F^2 \omega^2)^2}-\frac{12  \gamma^4 v_0^4 \pi^2}{5(\gamma^2 v_0^2+F^2 \omega^2)^2} r^2,
\end{equation}
which also indicates that in the limit of $r$ tending to zero $K<3$, implying that the stationary distribution is always platykurtic no matter how small $r$ is. This is also completely different from the results for pure harmonic potential. The numerical simulations are given in Fig. \ref{K}(b). The similar behavior for $\phi(\tau)=\frac{\alpha}{(1+\tau)^{1+\alpha}}$ can also be observed from Fig. \ref{K}(c), which still turns out to be platykurtic for any large $\alpha$.

\subsubsection{Relaxation dynamics}

\begin{figure}[htbp]
\centering
\includegraphics[scale=0.19]{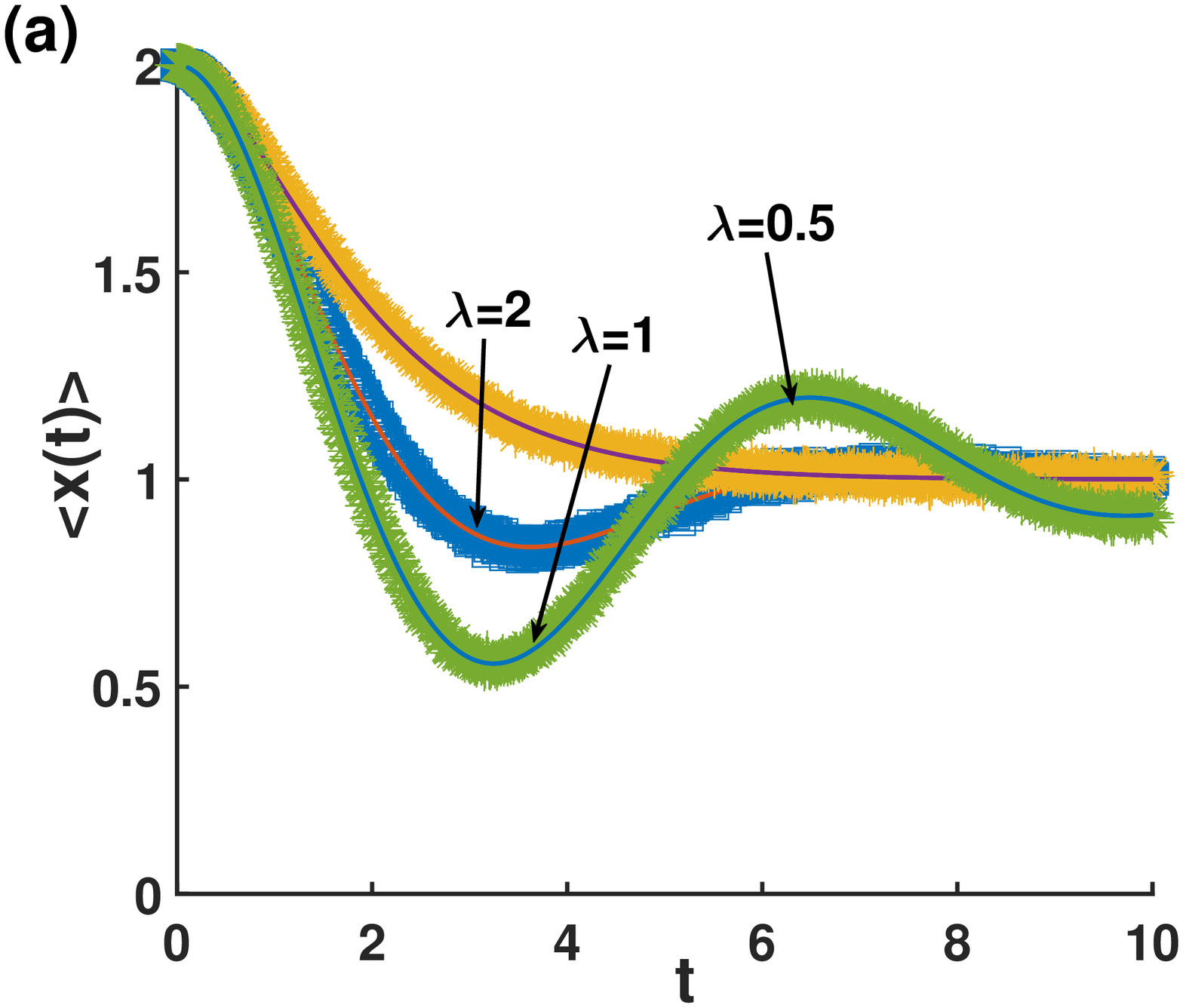}
\includegraphics[scale=0.19]{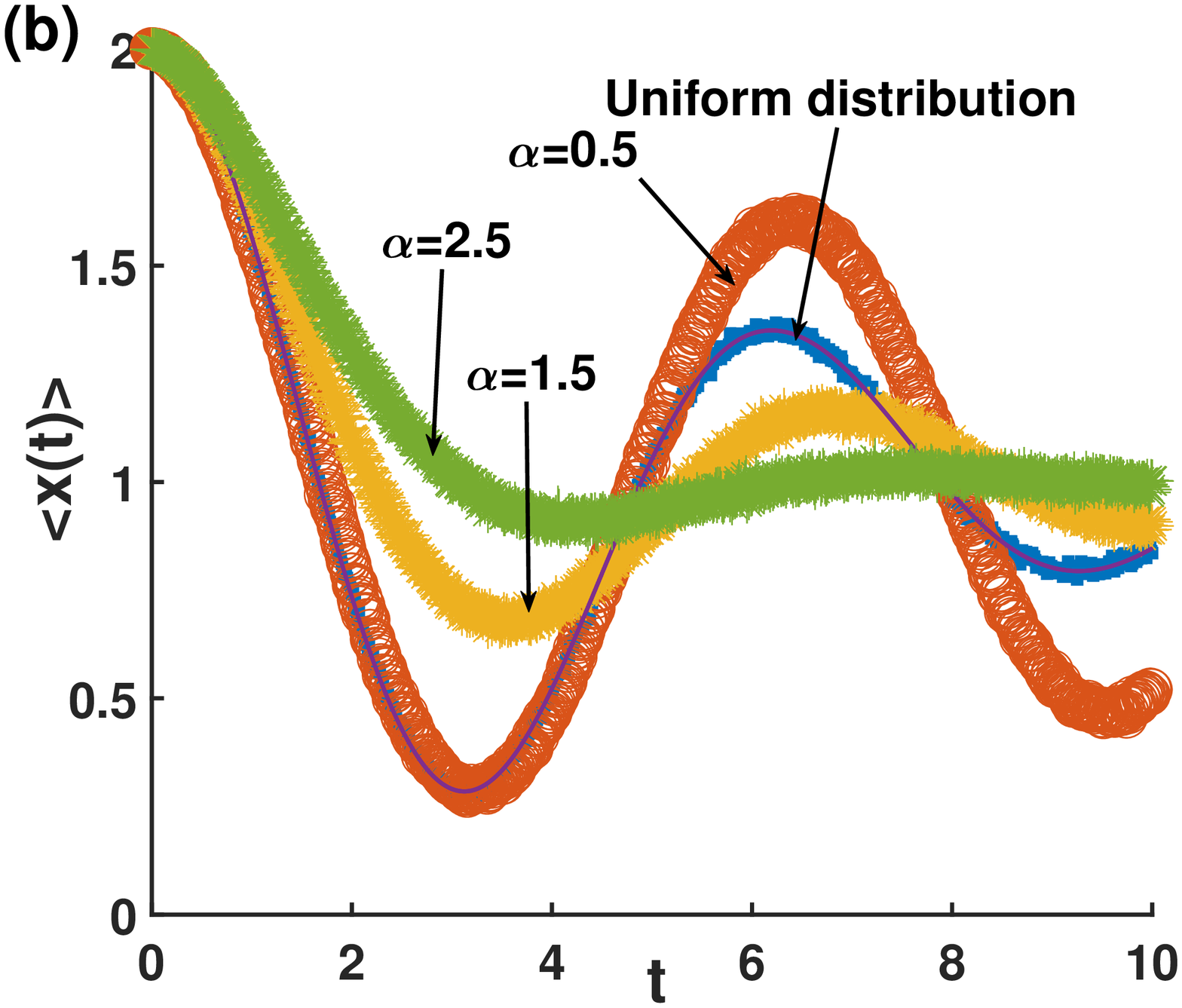}
\caption{Numerical simulations of average displacement of L\'{e}vy walk in constant force field combined with harmonic potential by sampling over $10^4$ realizations. The parameters are $v_0=1$ and $x_0=2$. For (a), we take $\phi(\tau)=\lambda e^{-\lambda\tau}$ with $\lambda=0.5$, $\lambda=1$, and $\lambda=2$; for (b), we take $\phi(\tau)$ to be uniform distribution on $[0,T]$ and power-law distribution with $\alpha=0.5, \alpha=1.5$, and $\alpha=2.5$. The solid lines in (a) are the theoretical results shown in \eqref{3.2}, and the solid lines in (b) are for the numerical inverse Laplace transform on \eqref{3.3}.}
\label{x02}
\end{figure}

In the last part of this section, we are going to discuss relaxation dynamics of L\'evy walk particle in mixed potential by assuming the initial position $x_0\neq 0$. In this case the iteration relation for $T_m(t)$ can be given by changing $H_m(0)$ in \eqref{2.11} to $H_m(x_0)$. Then for exponential distribution  $\phi(\tau)=\lambda e^{-\lambda\tau}$, by utilizing \eqref{a6} we have
\begin{equation*}
	\langle\hat{x}(s)\rangle=\sqrt{\pi}\widehat{R}_1(s)=\frac{F \omega^2+\gamma s (\lambda+s)x_0}{\gamma s (\lambda s+s^2+\omega^2)},
\end{equation*}
which has the following form after inverse Laplace transform
\begin{equation}\label{3.2}
\begin{split}
   \langle x(t)\rangle=& \frac{F}{\gamma}-\frac{e^{-\frac{1}{2}t(\lambda+\sqrt{\lambda^2-4 \omega^2})}(F-\gamma x_0)}{2\sqrt{\lambda^2-4 \omega^2}\gamma} \\
     &\times\left[\left(-1+e^{t\sqrt{\lambda^2-4 \omega^2}}\right)\lambda\right.\\
     &+\left.\left(1+e^{t\sqrt{\lambda^2-4 \omega^2}}\right)\sqrt{\lambda^2-4 \omega^2}\right].
\end{split}
\end{equation}
The result is verified by Fig. \ref{x02}(a). Besides from \eqref{3.2} we can also conclude that the relaxation is still exponential, and the value of initial velocity $v_0$ for each step has no influence on the average displacement, which are in accordance with the ones of \cite{Pengbo1}.

Next we will consider $\phi(\tau)=\omega/(2\pi) \mathbf{1}_{[0,2\pi/\omega]}(\tau)$, and the corresponding average displacement in Laplace space can be given as
\begin{equation}\label{3.3}
  \begin{split}
     \langle\hat{x}(s)&\rangle =\frac{1}{\gamma s(s^2+\omega^2)}\\
      &\times\frac{s\big(2 F \omega^2+\gamma(s^2-\omega^2)x_0\big)\cos(\omega T)}{e^{T s}(-s+s^2 T+T \omega^2)+s \cos(\omega T)-\omega \sin(\omega T)}\\
     &+\frac{1}{\gamma s(s^2+\omega^2)}\\
     &\times \frac{\omega \big(F (s^2-\omega^2)-2 \gamma x_0 s^2\big)\sin(\omega T)}{e^{T s}(-s+s^2 T+T \omega^2)+s \cos(\omega T)-\omega \sin(\omega T)} \\
       &+\frac{e^{T s}}{\gamma s(s^2+\omega^2)}\\
       &\times \frac{ F \omega^2(-2 s+s^2 T+T \omega^2)}{e^{T s}(-s+s^2 T+T \omega^2)+s \cos(\omega T)-\omega \sin(\omega T)}\\
       &+\frac{e^{T s}}{\gamma s(s^2+\omega^2)}\\
       &\times\frac{\gamma s (-s^2+s^3 T+\omega^2+s T \omega^2) x_0}{e^{T s}(-s+s^2 T+T \omega^2)+s \cos(\omega T)-\omega \sin(\omega T)}.
  \end{split}
\end{equation}
Figure \ref{x02}(b) presents the numerical inverse Laplace transform of \eqref{3.3} with $\phi(\tau)$ being power-law density. From Fig. \ref{x02} we can observe that the smaller $\lambda$ or $\alpha$ is, the initial decay will be faster and the oscillations will appear so it will take longer time for average displacement to  converge to position $F/\gamma$.

\section{Conclusion}\label{sec 4}

L\'evy walks, as spatiotemporally coupled random-walk processes describing superdiffusive, have been widely accepted and recognized.
In this paper, we mainly consider L\'evy walk in constant force field and mixed potential $V(x)=-Fx+\gamma^2/2 x^2$. For both of the potentials we establish models based on the dynamics of each step and the random walk theory. Hermite polynomials provide us the possibility to solve L\'evy walk process in external potential, which seems to be too hard to be treated by traditional integral transform method.

In the constant force field, the average displacements are calculated for different types of running times. Comparing with the free L\'evy walk process, for exponentially distributed waiting time the variance still linearly depends on time $t$ while the coefficient of it becomes bigger if $F\neq 0$, which indicates that the external force makes the process move faster. Besides for power-law distributed running time, the asymptotic forms of average displacement and MSD for long time do not rely on the initial velocity of each step $v_0$. From the numerical simulations of fractional moments, we discover that L\'evy process with $\phi(\tau)$ being power-law and $1<\alpha<2$ is still a strongly anomalous diffusion; what is different from free L\'evy walk in such case is that the turning point of the exponent of fractional moment $\langle |x(t)|^q\rangle$ locates at $q=\alpha/2$ instead of $\alpha$ and the exponents for regions $q<\alpha/2$ and $q>\alpha/2$ are also different.

For the mixed potential, the asymptotic behaviors of average displacement and MSD are also calculated; further from the result of variance we discover it is the same as the result obtained from pure harmonic potential. According to the trends of stationary distribution with different parameters, we can observe the procedure of bimodal changing to monomodal state; the additional force brings a translation and skewness to the stationary distributions comparing with the results of pure harmonic potential. Another major difference between pure potential and the mixed one is the kurtosis; the result of latter potential is always platykurtic if $F\neq 0$. In the final part, we discuss the relaxation dynamics; for exponentially distributed running time, it decays exponentially from the initial position $x_0$ to $F/\gamma$, and for some parameters the oscillations can be observed.

%

\section*{Acknowledgements}

This work was supported by the National Natural Science Foundation of China under Grant No. 12071195, and the AI and Big Data Funds under Grant No. 2019620005000775.

\begin{appendix}

\section{A brief introduction of Hermite polynomials}\label{Appen_A}

Hermite polynomials are a set of orthogonal polynomials defined on $(-\infty,\infty)$ with weight function $e^{-x^2}$ \cite{hermit_intro}. One way of standardizing the Hermite polynomials is given as
\begin{equation}\label{a1}
  H_n(x)=(-1)^n e^{x^2} \frac{d^n}{d x^n} e^{-x^2}.
\end{equation}
Furthermore, its orthogonality can be represented as
\begin{equation}\label{a2}
  \int_{-\infty}^{\infty} H_n(x)H_m(x)e^{-x^2}dx=\sqrt{\pi}2^n n! \delta_{n,m},
\end{equation}
where $\delta_{n,m}$ is the Kronecker delta function.
By Taylor's expansion, it has
\begin{equation}\label{a3}
  H_n(x+y)=\sum_{k=0}^{n}\binom{n}{k} H_k(y) (2 x)^{n-k},
\end{equation}
and the following holds
\begin{equation}\label{a4}
  H_n(\gamma x)=\sum_{j=0}^{\lfloor\frac{n}{2}\rfloor} \gamma^{n-2 j} (\gamma^2-1)^j \binom{n}{2 j} \frac{(2 j)!}{j!}H_{n-2 j}(x),
\end{equation}
where $\lfloor\frac{n}{2}\rfloor$ is the biggest integer smaller than $\frac{n}{2}$. The special value of Hermite polynomials is the value evaluated at zero argument $H_n(0)$, which are called Hermite number,
\begin{equation}\label{a5}
  H_n(0)=
  \begin{cases}
       0, &\mbox{if $n$ is odd} \\
       (-1)^{\frac{n}{2}} 2^{\frac{n}{2}} (n-1)!!, & \mbox{if $n$ is even}.
     \end{cases}
\end{equation}

In particular,
\begin{equation}\label{a6}
   H_0(x)=1,\quad H_1(x)=2 x,\quad  H_2(x)=4 x^2-2.
\end{equation}

\section{Derivations of \eqref{2.11} and \eqref{2.18}}\label{App_B}

\begin{widetext}
According to the property of Dirac delta-function, we can rewrite \eqref{2.2} and \eqref{2.3} as
\begin{align}
	   q(x_t,t)=& \frac{1}{2}\int_{0}^{t}\frac{\phi(\tau)}{|\cos(\omega\tau)|} q\left(\frac{F}{\gamma}+\frac{1}{\cos(\omega\tau)}\Big(x_t-\frac{F}{\gamma}-\frac{v_0}{\omega}\sin(\omega\tau)\Big),t-\tau\right)d\tau \nonumber\\
     &+\frac{1}{2}\int_{0}^{t}\frac{\phi(\tau)}{|\cos(\omega\tau)|} q\left(\frac{F}{\gamma}+\frac{1}{\cos(\omega\tau)}\Big(x_t-\frac{F}{\gamma}+\frac{v_0}{\omega}\sin(\omega\tau)\Big),t-\tau\right)d\tau +P_0(x)\delta(t),\label{2.4}\\
    P(x_t,t)=&\frac{1}{2}\int_{0}^{t}\frac{\Phi(\tau)}{|\cos(\omega\tau)|} q\left(\frac{F}{\gamma}+\frac{1}{\cos(\omega\tau)}\Big(x_t-\frac{F}{\gamma}-\frac{v_0}{\omega}\sin(\omega\tau)\Big),t-\tau\right)d\tau \nonumber \\
     &+\frac{1}{2}\int_{0}^{t} \frac{\Phi(\tau)}{|\cos(\omega\tau)|} q\left(\frac{F}{\gamma}+\frac{1}{\cos(\omega\tau)}\Big(x_t-\frac{F}{\gamma}+\frac{v_0}{\omega}\sin(\omega\tau)\Big),t-\tau\right)d\tau.\label{2.5}
\end{align}
Inserting the assumed form of $q$ in \eqref{1.10} into \eqref{2.4} results in
\begin{equation}\label{B3}
	\begin{split}
		\sum_{n=0}^{\infty} H_n(x)T_n(t)e^{-x^2}=\frac{1}{2}\int_0^t \frac{\phi(\tau)}{|\cos(\omega\tau)|}\sum_{n=0}^{\infty} \left[H_n(y_-)e^{-y_-^2}+H_n(y_+)e^{-y_+^2}\right]T_n(t-\tau) d\tau,
	\end{split}
\end{equation}
where $y_{\pm}=\frac{F}{\gamma}+\frac{1}{\cos(\omega\tau)}\Big(x_t-\frac{F}{\gamma}\pm\frac{v_0}{\omega}\sin(\omega\tau)\Big)$. Then multiplying $H_m(x)$ on both sides of \eqref{B3} and integrating w.r.t. $x$ over $(-\infty,\infty)$, after taking variable change we have
\begin{equation}\label{B4}
\begin{split}
	\sqrt{\pi}2^m m! T_m(t)&=\frac{1}{2}\int_0^t \frac{\phi(\tau)}{|\cos(\omega\tau)|}\sum_{n=0}^{\infty} \int_{-\infty}^{\infty}\left[H_n(y_-)e^{-y_-^2}+H_n(y_+)e^{-y_+^2}\right]H_m(x)dx\,T_n(t-\tau)d\tau\\
		&=\frac{1}{2}\int_0^t \phi(\tau) \sum_{n=0}^{\infty} \int_{-\infty}^{\infty} \left[H_m\left(\cos(\omega \tau) x -\frac{F}{\gamma} \cos(\omega\tau) +\frac{F}{\gamma} +\frac{v_0}{\omega} \sin(\omega\tau) \right)\right.\\
		&\quad\left.+H_m\left(\cos(\omega \tau) x -\frac{F}{\gamma} \cos(\omega\tau) +\frac{F}{\gamma} -\frac{v_0}{\omega} \sin(\omega\tau) \right) \right]H_n(x)e^{-x^2} dx\,T_n(t-\tau)d\tau,
\end{split}
\end{equation}
where the property \eqref{a2} is also utilized.  
From the properties of Hermite polynomials \eqref{a3} and \eqref{a4}, the integration in \eqref{B4} can be calculated as
\begin{equation*}
\begin{split}
	&\int_{-\infty}^{\infty} H_m\left(\cos(\omega \tau) x -\frac{F}{\gamma} \cos(\omega\tau) +\frac{F}{\gamma} +\frac{v_0}{\omega} \sin(\omega\tau) \right)H_n(x)e^{-x^2}dx\\
	&=\int_{-\infty}^{\infty} \sum_{l=0}^m \binom{m}{l} H_l\big(\cos(\omega \tau) x\big)\left(-\frac{2F}{\gamma} \cos(\omega\tau) +\frac{2F}{\gamma} +\frac{2v_0}{\omega} \sin(\omega\tau) \right)^{m-l}H_n(x)e^{-x^2}dx\\
	&= \sum_{l=0}^m \sum_{j=0}^{\lfloor\frac{l}{2}\rfloor}  \binom{m}{l} \binom{l}{2j} \frac{(2j)!}{j!} \cos^{l-2j}(\omega\tau)\left(\cos^2(\omega\tau)-1\right)^j \left(-\frac{2F}{\gamma} \cos(\omega\tau) +\frac{2F}{\gamma} +\frac{2v_0}{\omega} \sin(\omega\tau) \right)^{m-l}\\
	&\quad\times \int_{-\infty}^{\infty} H_{l-2j}(x)H_n(x)e^{-x^2}dx.
\end{split}
\end{equation*}
\end{widetext}
Then according to the orthogonal property of Hermite polynomial \eqref{a2}, we can finish calculating this part of integration. The similar method can give us another integration in the second equation of \eqref{B4}, and finally after some simplifications and applying Laplace transform w.r.t. $t$ we have \eqref{2.11}. On the other hand, from \eqref{2.5} and \eqref{1.11} after a similar derivation we can arrive at \eqref{2.18}.

\end{appendix}

\bibliography{ref}

\end{document}